\documentstyle[12pt]{article}
\def\al{\alpha}

\def\de{\delta}

\def\et{\eta}
\def\th{\theta}

\def\la{\lambda}

\def\si{\sigma}

\def\ph{\phi}

\def\ps{\psi}

\def\De{\Delta}
\def\Th{\Theta}

\def\fr#1#2{{{#1} \over {#2}}}

\def\frac#1#2{{\textstyle{{#1}\over {#2}}}}

\def\lsim{\mathrel{\rlap{\lower4pt\hbox{\hskip1pt$\sim$}}
    \raise1pt\hbox{$<$}}}
\def\gsim{\mathrel{\rlap{\lower4pt\hbox{\hskip1pt$\sim$}}
    \raise1pt\hbox{$>$}}}
\def\sqr#1#2{{\vcenter{\vbox{\hrule height.#2pt
         \hbox{\vrule width.#2pt height#1pt \kern#1pt
         \vrule width.#2pt}
         \hrule height.#2pt}}}}

\newcommand{\beq}{\begin{equation}}
\newcommand{\eeq}{\end{equation}}
\newcommand{\bea}{\begin{eqnarray}}
\newcommand{\eea}{\end{eqnarray}}
\newcommand{\rf}[1]{(\ref{#1})}

\renewenvironment{thebibliography}[1]
 { \rm
   \begin{list}{\arabic{enumi}.}
    {\usecounter{enumi} \setlength{\parsep}{0pt}
     \setlength{\itemsep}{3pt} \settowidth{\labelwidth}{#1.}
     \sloppy
    }}{\end{list}}

\begin{document}
\titlepage

\begin{flushright}
{COLBY-94-04\\}
{IUHET 277\\}
{quant-ph/9506009\\}
\end{flushright}
\vglue 1cm

\begin{center}
{{\bf  LONG-TERM EVOLUTION AND REVIVAL STRUCTURE\\
	OF RYDBERG WAVE PACKETS FOR HYDROGEN\\
	      AND ALKALI-METAL ATOMS
\\}
\vglue 1.0cm
{Robert Bluhm$^a$ and V. Alan Kosteleck\'y$^b$\\}
\bigskip
{\it $^a$Physics Department\\}
\medskip
{\it Colby College\\}
\medskip
{\it Waterville, ME 04901, U.S.A.\\}
\bigskip
{\it $^b$Physics Department\\}
\medskip
{\it Indiana University\\}
\medskip
{\it Bloomington, IN 47405, U.S.A.\\}

}
\vglue 0.8cm

\end{center}

{\rightskip=3pc\leftskip=3pc\noindent
This paper begins with
an examination of the revival structure and long-term
evolution of Rydberg wave packets for hydrogen.
We show that after the initial cycle of collapse
and fractional/full revivals,
which occurs on the time scale $t_{\rm rev}$,
a new sequence of revivals begins.
We find that the structure of the new revivals is
different from that of the fractional revivals.
The new revivals are characterized by periodicities
in the motion of the wave packet with periods
that are fractions of the revival time scale $t_{\rm rev}$.
These long-term periodicities result in the
autocorrelation function at times greater than $t_{\rm rev}$
having a self-similar resemblance to its structure
for times less than $t_{\rm rev}$.
The new sequence of revivals culminates with the formation
of a single wave packet that more closely resembles
the initial wave packet than does the full revival
at time $t_{\rm rev}$,
i.e., a superrevival forms.
Explicit examples of the superrevival structure
for both circular and radial wave packets are given.
We then study wave packets in alkali-metal atoms,
which are typically used in experiments.
The behavior of these packets is affected
by the presence of quantum defects that
modify the hydrogenic revival time scales and periodicities.
Their behavior can be treated analytically using
supersymmetry-based quantum-defect theory.
We illustrate our results for alkali-metal atoms
with explicit examples
of the revival structure for radial wave packets in rubidium.

}

\medskip
\centerline{\it To appear in the
June 1995 issue of Physical Review A}

\vfill
\newpage

\baselineskip=20pt

{\bf\noindent I. INTRODUCTION}
\vglue 0.4cm

A localized electron wave packet is formed when a short-pulsed
laser excites a coherent superposition of Rydberg states
\cite{ps,az}.
Wave packets of this type offer
the opportunity to investigate the classical limit
of the motion of electrons in Rydberg atoms.
Initially, the motion of the wave packet
follows the classical motion of a charged particle in
a Coulomb field.
The period of the motion is the classical period $T_{\rm cl}$
of a particle in a keplerian orbit.
However,
this motion persists only for a few cycles,
whereupon quantum interference effects cause the wave
packet first to collapse and then to undergo a sequence
of revivals
\cite{ps,az,ap,nau1}.
The revivals are characterized by the recombination
of the collapsed wave packet
into a form close to the original shape,
which again oscillates with period $T_{\rm cl}$.
The recombined wave packet is called a full revival,
and it appears at a time $t_{\rm rev}$.
For various times earlier than $t_{\rm rev}$,
the wave packet gathers into
a series of subsidiary wave packets
called fractional revivals.
The motion of these fractional revivals is periodic,
with a period equal to a rational fraction of the
classical keplerian period.

This cycle of collapse,
fractional revivals,
and full revival is characteristic of several
distinct types of Rydberg wave packets.
Among these are radial, circular, and elliptical packets.
A radial wave packet is localized in the radial coordinate
but is in a definite eigenstate of the angular momentum.
A circular wave packet is a superposition of fully
aligned eigenstates with maximal values of
the angular-momentum quantum numbers.
It is localized in both angular and radial coordinates, and
it follows a circular trajectory.
Lastly,
an elliptical wave packet is localized in all
three dimensions and travels along a classical
keplerian orbit.

Recently,
there has been much experimental interest in the study
of these Rydberg wave packets.
Radial wave packets can be formed by the excitation
of the atom
from the ground state by a short-pulsed laser.
The time evolution of the wave packet may then be
studied using a pump-probe method of detection
involving either time-delayed photoionization
\cite{az}
or phase modulation
\cite{noord,broers,christian}.
The periodic motion of the radial wave packet,
with the period equal to the classical orbital period,
has been observed
\cite{tenWolde,yeazell1},
and full and fractional revivals have been
seen experimentally
\cite{yeazell2,yeazell3,meacher}.
Circular states have been produced as well,
using either the adiabatic microwave transfer method
\cite{kleppner}
or crossed electric and magnetic fields
\cite{dg}.
These methods excite a range of states in a fixed
$n$ manifold,
resulting in a wave packet that is stationary.
A technique for the excitation of circular-orbital states
to generate a circular wave packet has been proposed
\cite{gns}.
If an additional weak electric field is present,
an elliptical wave packet of arbitrary eccentricity
could be produced.

Several theoretical approaches have been used to study
the properties of Rydberg wave packets.
Some involve a description via
different types of coherent states
\cite{klauder}.
Radial wave packets
have been studied both numerically
\cite{ps}
and perturbatively
\cite{az},
and recently a description as a type of squeezed
state has been obtained
\cite{rss,susywave}.
They exhibit oscillations in the
uncertainty product characteristic of a squeezed state.
These states undergo full and fractional revivals
with localized subsidiary waves distributed
along the radial direction.
In contrast,
the coherent-state approaches dealing with circular
and elliptical states
\cite{brown,mostowski,nieto,mca,nau2,gdb,gaeta}
all involve superpositions of angular-momentum eigenstates
and/or large angular-momentum quantum numbers.
The simplest such state is a circular wave packet
consisting of a superposition of aligned eigenstates
weighted by a gaussian function
\cite{gaeta}.
These have been studied numerically and exhibit both
fractional and full revivals.
The subsidiary waves that form in this case are at a
fixed radius and are distributed in the azimuthal angle.

Most work on the evolution of Rydberg wave packets
has focused on the structure of the fractional revivals,
which occur at times less than $t_{\rm rev}$.
In this paper,
we examine the evolution and revival structure
of Rydberg wave packets for times \it beyond \rm
the revival time.
We find that
after a few revival cycles the wave packet ceases
to reform at multiples of the revival time.
Instead, a new sequence of collapses and revivals begins
that is different from that of the usual fractional revivals.
This sequence culminates with the formation of a wave
packet that better resembles the original packet
than does the full revival at time $t_{\rm rev}$.
We refer to packets of this new type as superrevivals.

The appearance of the superrevivals is
controlled by a new time scale $t_{\rm sr}$.
We find that at certain times $t_{\rm frac}$
\it before \rm  $t_{\rm sr}$,
the wave function can be written as a sum of macroscopically
distinct subsidiary waves.
Their motion is periodic with period $T_{\rm frac}$
given in terms of $t_{\rm rev}$ and $T_{\rm cl}$.
This periodicity is on a much longer time scale than
that of the usual fractional revivals.
For times greater than $t_{\rm rev}$,
it produces in the autocorrelation function
a self-similar resemblance to its appearance
at times less than $t_{\rm rev}$.

In addition to presenting the above qualitative
features of the new revival structure,
this paper discusses a quantitative theory
providing detailed expressions for the
various time scales and periodicities that appear.
The features of the theory are illustrated with
explicit examples for both circular and
radial wave packets.
In these examples,
we first consider wave packets having
large values of the principle quantum number $n$,
since these display detailed features of the superrevivals
most strikingly.
We then consider wave packets with smaller values
of $n$ for which the time scales for the superrevival
structure fall within a range currently
accessible in experiment.

The above discussion applies to wave packets in hydrogen.
In the latter part of this paper,
we extend our analysis to obtain an analytical description
for the case of Rydberg wave packets in alkali-metal atoms,
typically examined in experiments.
Examples are given to illustrate the
resulting modifications of the behavior.
The issue of distinguishing effects on the revival times
and periodicities caused by the quantum defects
from those caused by a detuning of the laser
has been considered in ref.\ \cite{detunings}.
In fact, not only are the effects different,
but also the quantum-defect modifications to long-term
revival times cannot
be obtained by a direct scaling of the hydrogenic results.

The structure of this paper is as follows.
Sec.\ II presents the generic form we use for
circular and radial wave packets
and introduces the distinct times scales
$T_{\rm cl}$, $t_{\rm rev}$, and $t_{\rm sr}$
determining wave-packet behavior
over the time period of interest.
In Sec.\ III,
we examine the long-term behavior of Rydberg wave
packets for hydrogen.
We show that at certain times $t_{\rm frac}$
the wave function is well approximated as
a sum over subsidiary waves with specified coefficients,
and we present constraints on the allowed
values of $t_{\rm frac}$.
We also demonstrate that at these times
the motion of the
subsidiary wave packets is periodic with a period
$T_{\rm frac}$ dependent on $t_{\rm rev}$ and $T_{\rm cl}$.
The periodicity of the autocorrelation function
is analyzed.
Sec.\ IV
provides illustrative examples of circular and radial
wave packets in hydrogen both for large values
of $n$ and for smaller values that are experimentally viable.

In Sec.\ V,
we extend our results to the
case of wave packets in alkali-metal atoms,
establishing the modifications in the
revival time scales and periodicities
that result from the inclusion of quantum defects.
Explicit examples for the
case of rubidium are provided in Sec.\ VI.
Section VII discusses the results
and considers some of the issues involved
in experimental verification of our theory.
Throughout the paper, technical details of the various proofs are
relegated to appendices.

\vglue 0.6cm
{\bf\noindent II. TIME SCALES FOR WAVE PACKET EVOLUTION}
\vglue 0.4cm

This section establishes some notation and conventions
and provides a brief review of the evolution
of Rydberg wave packets in hydrogen for
times up to the revival time $t_{\rm rev}$.
In Sec.\ IIA,
we present the wave function
we use throughout the paper for the analysis.
It is a generic form that permits
simultaneous treatment of circular and radial
wave packets.
In Sec.\ IIB,
we expand the component eigenenergies in a Taylor series
about the wave-packet energy.
This leads to the definition
of distinct time scales governing the
evolution of the wave function.
Some background on the behavior of hydrogenic wave packets
for times up to $t_{\rm rev}$ is given
in Sec.\ IIC,
along with an example.

\vglue 0.6cm
{\bf\noindent A. Circular and Radial Wave Packets}
\vglue 0.4cm

The time-dependent wave function for a hydrogenic
wave packet may be expanded in terms of energy eigenstates as
\beq
\Psi ({\vec r},t) = \sum_{nlm} c_{nlm}
\psi_{nlm}({\vec r}) \exp \left[ -i E_n t \right]
\quad .
\label{wave1}
\eeq
Here,
$E_n = -1/2 n^2$ is the energy in atomic units,
and $\psi_{nlm}({\vec r})$ is a
hydrogen eigenstate of
energy and angular momentum.
The weighting coefficients $c_{nlm} = \left<
\psi_{nlm}({\vec r}) \vert \Psi ({\vec r},0) \right>$
depend on the initial wave function.

Two types of wave packets
used in experiments
are considered in this paper.
The first type is the
\it circular \rm wave packets,
which consist of a sum of
fully aligned eigenstates,
i.e.,
ones with $l=m=n-1$.
In this case,
the sum in
Eq.\ \rf{wave1}
reduces to a sum over $n$ alone,
and the weighting coefficients $c_n$
are independent of $l$ and $m$.
The second type is the
\it radial \rm wave packets,
which consist of a sum of eigenstates
with fixed angular-momentum quantum number.
For example,
all the eigenstates may be p states.
Once again,
the sum in Eq.\ \rf{wave1} depends only on $n$,
and we may drop the $l$ and $m$ subscripts on $c_n$.

Adopting the generic notation that $\varphi_n ({\vec r})$
represents the eigenstates appropriate to a given type
of wave packet,
we may then rewrite
Eq.\ \rf{wave1} as
\beq
\Psi ({\vec r},t) = \sum_{n=1}^{\infty} c_n
\varphi_n ({\vec r}) \exp \left[ -i E_n t \right]
\quad .
\label{wave2}
\eeq
For the circular wave packets
$\varphi_n ({\vec r}) = \psi_{n,n-1,n-1}({\vec r})$,
while for the radial wave packets
$\varphi_n ({\vec r}) = \psi_{n,1,0}({\vec r})$.

Both types of wave packet are
excited by a short laser pulse.
Since the laser can be tuned to excite coherently
a superposition of states centered on a mean value
of the principal quantum number $\bar n$,
in what follows we assume that the distribution is strongly
centered around a value $\bar n$.
We may therefore approximate the square of the weighting
coefficients as a gaussian function
\beq
\bigm| c_n \bigm|^2 = \fr 1 {\sqrt{2 \pi \si^2}}
e^{- \fr {(n - {\bar n})^2} {2 \si^2}}
\quad .
\label{cn}
\eeq
In the derivations that follow,
we take $\bar n$ to be an integer.
The results can readily be extended to incorporate
noninteger $\bar n$.
The effects of noninteger $\bar n$
and the consequences of laser detuning
are examined in
ref.\ \cite{detunings}.

\vglue 0.6cm
{\bf\noindent B. Time Scales}
\vglue 0.4cm

If we expand the energy in a Taylor series around the centrally
excited value $\bar n$,
we obtain
\beq
E_n \simeq E_{\bar n} + E_{\bar n}^\prime (n - {\bar n})
+ \fr 1 2 E_{\bar n}^{\prime\prime} (n - {\bar n})^2
+ \fr 1 6 E_{\bar n}^{\prime\prime\prime} (n - {\bar n})^3
+ \cdots
\quad ,
\label{energy}
\eeq
where each prime on $E_{\bar n}$
denotes a derivative.
The terms with derivatives define distinct time scales that depend
on $\bar n$.
The first time scale,
\beq
T_{\rm cl} = \fr {2 \pi} {E_{\bar n}^\prime} = 2 \pi {\bar n}^3
\quad ,
\label{tcl}
\eeq
is the classical keplerian period.
It controls the initial behavior of the packet.
The second time scale,
\beq
t_{\rm rev} = \fr {- 2 \pi} {\fr 1 2 E_{\bar n}^{\prime\prime}}
= \fr {2 {\bar n}} 3 T_{\rm cl}
\quad ,
\label{trev}
\eeq
is the revival time.
It governs the appearance of
the usual fractional and full revivals.

The subject of this paper is the behavior of
the packet on time scales greater than $t_{\rm rev}$.
This behavior is dominantly controlled by a third time scale,
\beq
t_{\rm sr} = \fr {2 \pi} {\fr 1 6 E_{\bar n}^{\prime\prime\prime}}
= \fr {3 {\bar n}} 4 t_{\rm rev}
\quad .
\label{tsr}
\eeq
The scale $t_{\rm sr} \gg t_{\rm rev}$
is a larger time scale we refer to as the
superrevival time.
Note that although $t_{\rm sr}$ is large
(typically about two orders of magnitude greater
than $t_{\rm rev}$)
for the range of $\bar n$ of interest,
it is still much smaller than the lifetime of
the excited Rydberg atom.
It is also smaller than the typical time scales
for which microwave black-body radiation causes
transitions
\cite{black-body}.
It is therefore experimentally and
theoretically reasonable to examine the behavior of the
packet on this time scale.

Keeping terms through third order,
and defining the integer index $k=n-{\bar n}$, we may rewrite
Eq.\ \rf{wave2} as
\beq
\Psi ({\vec r},t) = \sum_{k=-\infty}^{\infty} c_k
\varphi_k ({\vec r}) \exp \left[ -2 \pi i
\left( \fr {kt} {T_{\rm cl}} -  \fr {k^2 t} {t_{\rm rev}}
+ \fr {k^3 t} {t_{\rm sr}} \right) \right]
\quad .
\label{psi3rd}
\eeq
We assume that $\bar n$ is large so that the lower limit
in the sum may be approximated by $- \infty$.
In this notation $c_k$ and $\varphi_k ({\vec r})$
represent $c_{n=k+{\bar n}}$ and
$\varphi_{n=k+{\bar n}} ({\vec r})$,
respectively.

\vglue 0.6cm
{\bf\noindent C. Review of Behavior within the Revival Time Scale}
\vglue 0.4cm

This subsection briefly summarizes the known behavior
of the packet within the revival time scale
and provides an example useful for comparison
in our later analysis.

The quantum evolution of the wave function \rf{wave2}
up to times of order $t_{\rm rev}$
is governed by the first two terms
in the exponential function in
Eq.\ \rf{psi3rd}.
If only the term linear in $k$ were present,
the wave packet would evolve like any localized
packet for the harmonic oscillator,
following the classical motion and
oscillating indefinitely with period $T_{\rm cl}$.

The higher-order terms modify this behavior.
The term quadratic in $k$ governs
the collapse and revival of the packet
for times within $t_{\rm rev}$.
An analysis of the role of this term
is given in ref.\ \cite{ap},
where it is shown that
at the times $t \approx \fr m n t_{\rm rev}$,
where $m$ and $n$ are relatively prime integers,
the wave packet may be rewritten approximately as
an equally weighted sum of subsidiary wave packets.
These are the usual fractional revivals.

The behavior of the
fractional revivals may be studied
by examining the absolute square of the autocorrelation function
\cite{nau1},
\beq
\vert A(t) \vert^2 = {\Bigm| \sum_n \vert c_n
\vert^2 e^{-i E_{n} t} \Bigm|}^2
\quad .
\label{auto}
\eeq
Since the wave function $\varphi_{n} ({\vec r})$
has been integrated over,
we find the same autocorrelation function
and hence the same revival structure for
both circular and radial wave packets.
At the fractional-revival times
$t \approx \fr m n t_{\rm rev}$,
the autocorrelation function is periodic with
fractional periods $\fr 1 r T_{\rm cl}$,
where $r$ denotes the number of subsidiary wave packets.

An example useful for comparison with our analyses below
is a wave packet in hydrogen centered about ${\bar n} = 319$
with $\si = 2.5$.
Figure 1 shows the square of the autocorrelation
function for times up to and just beyond
$t_{\rm rev} \simeq 1.05$ $\mu$sec.
Fractional revivals are prominent at
$t \approx \fr 1 8 t_{\rm rev} \simeq 0.13$ $\mu$sec,
$t \approx \fr 1 6 t_{\rm rev} \simeq 0.18$ $\mu$sec,
$t \approx \fr 1 4 t_{\rm rev} \simeq 0.26$ $\mu$sec,
$t \approx \fr 1 2 t_{\rm rev} \simeq 0.53$ $\mu$sec,
and at multiples of these values.
The periodicity of the autocorrelation function
at the one-half and full revivals is
$T_{\rm cl} \simeq 4.9$ nsec.
This can be seen in the figure.
The periodicities at the earlier fractional revivals,
for $t \approx \fr 1 8 t_{\rm rev}$,
$\fr 1 6 t_{\rm rev}$,
$\fr 1 4 t_{\rm rev}$,
respectively,
are
$\fr 1 4 T_{\rm cl}$,
$\fr 1 3 T_{\rm cl}$,
$\fr 1 2 T_{\rm cl}$.
As can be observed in the figure,
the peaks in the autocorrelation function are not
as pronounced for times approaching $t_{\rm rev}$
as they are for earlier times.
This is because higher-order terms in the
energy expansion \rf{energy} act to distort the revivals.

\vglue 0.6cm
{\bf\noindent III. WAVE PACKETS IN HYDROGEN}
\vglue 0.4cm

In this section,
we show that at certain times $t_{\rm frac}$
it is possible to expand
the third-order wave function $\Psi ({\vec r},t)$
of Eq.\ \rf{psi3rd} as a series of subsidiary wave functions.
Following
ref.\ \cite{ap},
the idea is to express
$\Psi ({\vec r},t)$ as a sum of
wave functions $\ps_{\rm cl}$
with matching periodicities
and with shape similar to that of
the initial wave function $\Psi ({\vec r},0)$.
The form of the coefficients in the
expansion of $\Psi ({\vec r},t)$ is obtained and
used to show that at times $t_{\rm frac}$ the expansion
correctly reproduces the wave function.
At particular values of $t_{\rm frac}$,
superrevivals are shown to form that more closely resembles
the initial wave packet than does the full revival.
Lastly,
we show that the wave function and
autocorrelation function are periodic with a
given period $T_{\rm frac}$.

In what follows,
we find it is useful to write the times
$t_{\rm frac}$ at which the subsidiary
wave packets form as a linear combination of rational
fractions of $t_{\rm sr}$ and $t_{\rm rev}$.
We therefore define
\beq
t_{\rm frac} = \fr p q t_{\rm sr} - \fr m n t_{\rm rev}
\quad ,
\label{times}
\eeq
where $p$ and $q$
are relatively prime integers,
as are $m$ and $n$.
Constraints on the possible values of these integers
are derived below.
In particular,
we find that $m$ is nonzero whenever $\bar n$
cannot be evenly divided by four.
We therefore write $\bar n$ as
\beq
{\bar n} = 4 \et + \la
\quad ,
\label{nbar}
\eeq
where $\et$ and $\la$ are both integers,
and $\la = 0$, $1$, $2$, or $3$.

The subsidiary wave functions we use in the expansion
of the full wave function are defined in terms of
a function $\ps_{\rm cl}$ depending only on the
first-derivative term in the energy expansion,
\beq
\psi_{\rm cl} ({\vec r},t) = \sum_{k=-\infty}^{\infty} c_k
\varphi_k ({\vec r}) \exp \left[ -2 \pi i
\fr {2 {\bar n}} 3 \fr {kt} {t_{\rm rev}} \right]
\quad .
\label{psicl}
\eeq
We write it in terms of $t_{\rm rev}$ instead of
$T_{\rm cl}$ because the former is more useful
when considering times greater than $t_{\rm rev}$.

The subscript on $\psi_{\rm cl}$
is a reminder that this wave function follows the classical motion
at all times.
Note also that at time zero,
$\psi_{\rm cl} ({\vec r},0) = \Psi ({\vec r},0)$,
i.e.,
$\psi_{\rm cl} ({\vec r},0)$
matches the initial wave function exactly.
These features of the function
$\psi_{\rm cl} ({\vec r},t)$
make it a suitable candidate for use in
generating an expansion of the full wave function
$\Psi ({\vec r},t)$
as a sum of macroscopically distinct subsidiary wave functions.

Appendix A shows that at the times $t_{\rm frac}$ the
wave function $\Psi ({\vec r},t)$ can be written
as a sum of subsidiary waves having the form of
$\psi_{\rm cl} ({\vec r},t)$,
but at times that are shifted by a fraction of the
period $T_{\rm cl}$.
The shifted functions have the form
$\psi_{\rm cl} ({\vec r},t + \fr {s \al} l T_{\rm cl})$,
where $l$ and $\al$ are certain specified integers
and $s = 0$, $1$, $2$, $\ldots$, $l-1$.
Since these functions form a set with the same
periodicity as $\Psi ({\vec r},t)$ at the times $t_{\rm frac}$,
we can use them as a basis for expanding $\Psi ({\vec r},t)$.
Moreover,
the shapes of the shifted $\psi_{\rm cl}$
resemble the initial wave function
at different points in its cycle,
after it has been time-translated
by a fraction of the classical period $T_{\rm cl}$.
An expansion of this type provides
a natural formalism within which to examine
the occurrence of long-term revivals.

We therefore write an expansion in subsidiary wave functions
at times $t \approx t_{\rm frac}$ in the form
\beq
\Psi ({\vec r},t) = \sum_{s=0}^{l-1} b_s
\psi_{\rm cl} ({\vec r},t + \fr {s \al} l T_{\rm cl} )
\quad .
\label{expans}
\eeq
The coefficients $b_s$ are complex valued and are given as
\beq
b_s = \fr 1 l \sum_{k^\prime =0}^{l-1}
\exp \left[ 2 \pi i \fr {\al s} l k^\prime \right]
\exp \left[ 2 \pi i \th_{k^\prime} \right]
\quad ,
\label{bs}
\eeq
where $\th_{k}$ is defined in
Eq.\ \rf{etatheta} of Appendix A.

Appendix A also derives the constraints
on the integers $p$, $q$, $m$, and $n$
in the definition \rf{times} of $t_{\rm frac}$,
and shows that $p=1$,
$q$ is a multiple of three,
and $m/n$ obeys Eq.\ \rf{mn}.
With these constraints,
the expansion of $\Psi ({\vec r},t)$ in terms
of subsidiary waves $\psi_{\rm cl} ({\vec r},t)$
holds at times
$t_{\rm frac}$ given by Eq.\ \rf{times}.
Since $t_{\rm sr} \gg t_{\rm rev}$,
we conclude that interesting behavior of the
wave packet can be expected
near times that are simple fractions of $\fr 1 3 t_{\rm sr}$.
This somewhat counterintuitive result
is strongly supported by the examples presented
in Sec.\ IV.

Whereas the usual fractional revivals
consist of $r$ subsidiary packets weighted
equally by factors of $1/r$,
and hence have fractional period $T = T_{\rm cl}/r$,
we find the superrevivals have different behavior.
An interesting feature of the $b_s$ coefficients
\rf{bs} at
the times $t_{\rm frac}$ is that they may have
\it different \rm moduli $\vert b_s \vert^2$,
and hence distinct subsidiary wave packets
may not all have the same weight.
Explicit examples of this are given in Sec.\ IV.
In certain cases,
all the $b_s$ coefficients except one
vanish.
The remaining one then has modulus one,
corresponding to the formation of one wave packet.
It turns out that at the times $t_{\rm frac}$,
even if the subsidiary wave packets are unequally weighted,
the wave packet and the autocorrelation function are periodic.

In what follows,
it is useful to introduce terminology
distinguishing the different types
of superrevival that can appear.
We refer to the set of subsidiary packets
at the times $t_{\rm frac}$
as a {\it fractional} superrevival.
However,
occasionally only a single packet appears,
resembling the initial wave packet more closely than the
usual full revival does at the time $t_{\rm rev}$.
This is because contributions are incorporated from
higher-order corrections in the expansion of
the time-dependent phase.
We call a single wave packet of this type a {\it full} superrevival.

After the formation of a fractional superrevival
consisting of several subsidiary wave packets,
it often happens that the subsidiary packets quickly
evolve into a configuration where one of them
is much larger than the others.
The dominant wave packet in this case can again resemble
the initial wave packet more closely than does the full
revival at time $t_{\rm rev}$.
We refer to a configuration of this type,
consisting of primarily one large wave packet,
as a {\it partial} superrevival.

Appendix B proves that the wave packet
$\vert \Psi ({\vec r},t) \vert^2$
as well as the absolute square of the autocorrelation function
$A(t_{\rm frac}) = \big< \Psi ({\vec r},0)
\vert \Psi ({\vec r},t_{\rm frac}) \big>$
are periodic at times $t \approx t_{\rm frac}$.
The period $T_{\rm frac}$ is given as
\beq
T_{\rm frac} = \fr 3 q t_{\rm rev} - \fr u v T_{\rm cl}
\quad ,
\label{Tfrac}
\eeq
where the integers $u$ and $v$ satisfy
Eq.\ \rf{uv2}.
In addition,
the value of $p$ in $t_{\rm frac}$
must equal $1$ for this periodic behavior to occur.

This periodicity is different from that
of the usual fractional revivals,
for which the periods are fractions of $T_{\rm cl}$.
Instead,
the periods of the fractional superrevivals
are fractions of $t_{\rm rev}$ combined
with small additional shifts that depend on $T_{\rm cl}$.
The time scale is considerably greater for the
fractional superrevivals than for the fractional revivals.
In addition,
this periodicity causes the autocorrelation function for
times much greater than $t_{\rm rev}$ to have
a self-similar resemblance to its form for
times less than $t_{\rm rev}$.
We will illustate this behavior explicitly
in the next section.

\vglue 0.6cm
{\bf\noindent IV. EXAMPLES FOR HYDROGEN}
\vglue 0.4cm

This section
presents several examples illustrating
in detail the various aspects
of the long-term evolution of Rydberg wave packets
in hydrogen.
Results numerically generated from analytical expressions
are compared
with the theoretical predictions given above.

In these examples,
we assume the laser is tuned to excite coherently
a distribution strongly centered at
either $\bar n = 319$ or $\bar n = 36$.
The former is large
and ensures that interesting features of the
the superrevival structure are noticeable.
Since 319 is not divisible by four,
the quantity $\la$ in Eq.\ \rf{nbar} is nonzero,
which in turn makes the ratio $m/n$ nonzero in the
definition of $t_{\rm frac}$.
The full analysis of Sec.\ III is therefore required.
An example with ${\bar n} = 320$,
which is divisible by four,
can be found in ref.\ \cite{sr}.

The other choice,
$\bar n = 36$,
is motivated by experimental considerations.
For this case,
the associated time delay in a pump-probe
experiment falls within a range that is currently accessible.
Our analysis shows that the
corresponding full superrevival occurs after
approximately 776 psec,
which is within the range of time delays considered in
refs.\ \cite{yeazell2,yeazell3}.

Sec.\ IVA
examines the autocorrelation functions for the
example with ${\bar n} = 319$.
Since the autocorrelation functions for corresponding
circular and radial wave packets are identical,
this provides a relatively broad perspective
on the long-term behavior.
In Sec.\ IVB,
we examine detailed features of circular wave packets
with $\bar n = 319$ at the
times $t_{\rm frac}$.
Hydrogenic radial wave packets with
the experimentally accessible value ${\bar n}=36$
are considered in Sec.\ IVC.

\vglue 0.6cm
{\bf\noindent A. Autocorrelation Functions}
\vglue 0.4cm

This subsection
discusses the autocorrelation functions for our
two choices of $\bar n$.
The absolute squares of the autocorrelation
functions are computed directly using the analytical expression in
Eq.\ \rf{auto}
and the definition of the $c_n$ coefficients given in
Eq.\ \rf{cn}.
These results are then compared with the preceding theoretical
analysis.
In what follows,
we restrict ourselves to times of order
$\fr 1 6 t_{\rm sr}$,
corresponding to a minimum value of $q$ equal to six.
An accurate treatment on longer time scales
requires fourth-order and higher
terms in the Taylor-series expansion of the energy.

The autocorrelation function is primarily useful in
the evaluation of theoretical predictions
because it is sensitive to the formation of single packets.
Thus,
in the following examples both full and partial superrevivals
appear as distinct peaks.
However,
fractional ones are less apparent.
For this reason,
we defer
to Secs.\ IVB and IVC
comparison of the results of numerical computations
with any details of our theory
involving, for instance, the
explicit values of the coefficients $b_s$.

As an example,
consider the autocorrelation function for a
Rydberg wave packet in hydrogen with ${\bar n} = 319$
and $\si = 2.5$.
In this case,
$\et = 79$, $\la = 3$, and $t_{\rm sr} \simeq 251$
$\mu$sec.
The absolute square of the autocorrelation function
is shown in Fig.\ 2.
Figure 2a shows the behavior of the wave packet for
the first 15 $\mu$sec.
The first $1.2$ $\mu$sec contains the cycle of
fractional and full revivals shown in
Fig.\ 1.
This cycle collapses after
approximately 5 $\mu$sec,
whereupon new structure emerges.
Figures 2b and 2c show $\vert A(t) \vert^2$ for times
up to and just beyond $\fr 1 6 t_{\rm sr}$.

Our prediction for $q=6$ is that a single wave packet should
form at $t \approx t_{\rm frac} \simeq 41.4$ $\mu$sec.
Only one wave packet appears because
only one $b_s$ coefficient is nonzero
(see Sec.\ VIB).
Equation \rf{Tfrac} predicts
the period of this packet
as $T_{\rm frac} \simeq 0.52$ $\mu$sec.
In Fig.\ 2c,
we observe large peaks in $\vert A(t) \vert^2$ near
42 $\mu$sec,
with a periodicity approximately equal to $T_{\rm frac}$.
These peaks are larger than the full-revival peak at
$t \approx t_{\rm rev} \simeq 1.05$ $\mu$sec.
This is because at $t \simeq 41.4$ $\mu$sec
the higher-order corrections to the time-dependent
phase are contributing coherently to the wave function.
Note also that between the large peaks
at $t \simeq 41.4$ $\mu$sec
there are smaller peaks
with periodicities that are fractions of
$T_{\rm frac}$.
These correspond to the formation of fractional
revivals during the times between the periodic
appearances of the full superrevival.

For $q=9$,
the theory gives $t_{\rm frac} \simeq 27.6$ $\mu$sec and
$T_{\rm frac} \simeq 0.35$ $\mu$sec.
Fig.\ 2b shows that the autocorrelation function
exhibits peaks near this time,
but with approximately half the period $T_{\rm frac}$.
This occurs because more than one subsidiary wave packet is forming,
and they are unequally weighted.
In this case,
the fractional superrevival evolves into a partial
superrevival at two separate times during each cycle.

The $q=12$ peaks at $t_{\rm frac} \simeq 20.7$ $\mu$sec
with $T_{\rm frac} \simeq 0.26$ $\mu$sec are more
pronounced than the $q=9$ peaks.
This is because
the corresponding subsidiary wave packets
evolve into a structure with
a single dominant wave packet,
thereby creating large peaks
in the autocorrelation function.
The periodicity of the peaks
agrees with the predicted value of $T_{\rm frac}$.

No prominent peaks appear in Fig.\ 2b
for $q=15$ at the times $t_{\rm frac} \simeq 16.6$ $\mu$sec.
As we show in the next subsection,
in this case there are four unequally
weighted subsidiary wave packets.
These evidently do not evolve into a structure
with one dominant wave packet.

Since $q=18$ is divisible by 9,
the number of terms in the expansion of the wave
function in this case is $l = \fr q 3 = 6$.
Of these six terms,
three vanish.
The remaining terms are unequally weighted with
one of them substantially larger than the rest.
This leads to the formation of a single packet
at time $t_{\rm frac} \simeq 13.8$ $\mu$sec,
with a period $T_{\rm frac} \simeq 0.17$ $\mu$sec.
Figure 2a shows peaks in the autocorrelation function
at this time with a periodicity that agrees with
$T_{\rm frac}$.

Peaks also appear in the autocorrelation function
near $t \simeq 7$ $\mu$sec.
For $q=36$,
the theory gives $t_{\rm frac} \simeq 6.91$ $\mu$sec
and $T_{\rm frac} \simeq 0.09$ $\mu$sec,
which agrees with the observed periodicity.
In this case,
however,
we find that there are 12 nonvanishing $b_s$
coefficients that are unequal in magnitude.
It is unclear why these subsidiary waves
evolve into a partial superrevival.

\vglue 0.6cm
{\bf\noindent B. Circular Wave Packets}
\vglue 0.4cm

This subsection examines more closely our
theoretical description of circular Rydberg wave packets
in hydrogen with ${\bar n} = 319$ and $\si = 2.5$.
The time-dependent wave function is given analytically in
Eq.\ \rf{wave2}
with the $c_n$ coefficients specified by
Eq.\ \rf{cn}
and with
$\varphi_n ({\vec r}) = \psi_{n,n-1,n-1}({\vec r})$.
Plots of the wave packet at different times
can be obtained numerically by taking advantage
of the peaking of the
$c_n$ coefficients around the value $\bar n$.
Figure 3 shows the azimuthal dependence at various times
of a cross-sectional slice of the wave packet
taken in the plane of the orbit and at fixed radius equal to
the expectation value $\left< r \right> = \fr 1 2
{\bar n} \left(2 {\bar n} + 1 \right)$.

The initial wave packet at time $t=0$ is
presented in Figure 3a.
It is localized about the azimuthal angle $\ph = 0$.
Figure 3b shows the first full revival,
at time $t_{\rm rev}$.
At this point,
the packet has collapsed,
passed through the sequence of fractional revivals,
and reformed.
It is primarily a single packet
shifted in phase relative to the initial one,
but some smaller subsidiary packets
remain visible.

The remainder of the graphs in Fig.\ 3
show the wave packet at times relevant for
comparison with our theoretical analysis.
In what follows,
we give the $b_s$ coefficients for the expansion
\rf{expans}
of the wave packet at the times $t_{\rm frac}$
for each of the cases $q=6$, $9$, $12$, $15$, and $18$.
We also compare the packets at the times
$t_{\rm frac}$ and $t_{\rm frac} + T_{\rm frac}$
to test the predicted periodicity.
For ${\bar n} = 319$,
we have $\eta = 79$ and $\la = 3$.
The value of $l$ is given in
Eq.\ \rf{lvalues}
and depends on $q$.
The integer $\al$ that appears in the expansion
of $\Psi ({\vec r},t)$ in
Eq.\ \rf{expans}
is equal to ${8 \et}/N$ where N is the product
of all the factors of $(8 \et)$ that are also factors of $l$.
Note that we set the integer $a$ in Eq.\ \rf{reduce}
equal to 1.
In fact, its value is irrelevant:
if it changes,
the $b_s$ coefficients become permuted
but leave unaffected the total sum of the subsidiary waves.

For $q=6$,
we obtain $l=6$, $N=8$, $\al = 79$,
$t_{\rm frac} = \fr 1 {6} t_{\rm sr} - \fr 3 8 t_{\rm rev}$,
and $T_{\rm frac} = \fr 1 {2} t_{\rm rev} - \fr 5 6 T_{\rm cl}$.
Using Eq.\ \rf{bs},
we find that $b_4 = 1$,
while all the other $b_s$
coefficients vanish.
The wave function at $t_{\rm frac}$ can therefore
be written as a single subsidiary packet.
Since $\psi_{\rm cl}$ is periodic with period $T_{\rm cl}$,
it is sufficient to evaluate the phase of
$\psi_{\rm cl} (t_{\rm frac} + \fr {s \al} l T_{\rm cl})$
for $s=4$
modulo $T_{\rm cl}$.
Suppressing the $\vec r$ dependence,
we find
$\Psi (t_{\rm frac}) \approx \psi_{\rm cl} (0)$.
Figure 3c shows the wave packet at the time $t_{\rm frac}$.
It does indeed consist of a single wave packet at the
initial point in the orbital cycle.
Comparing Fig.\ 3c to Fig.\ 3b,
which represents the wave packet at the full revival,
we see that the full superrevival resembles the initial
wave in Fig.\ 1a more closely than does the full revival
at $t=t_{\rm rev}$.
Moreover,
the tail of subsidiary waves visible in Fig.\ 3b
is absent in Fig.\ 3c.
The wave packet at the time
$t_{\rm frac} + T_{\rm frac}$
is plotted in Fig.\ 3d.
This figure resembles Fig.\ 3c,
which verifies the predicted periodicity of the wave packet.

With $q=9$,
we obtain $l=3$, $N=1$, $\al = 632$,
$t_{\rm frac} = \fr 1 {9} t_{\rm sr} - \fr 1 4 t_{\rm rev}$,
and $T_{\rm frac} = \fr 1 {3} t_{\rm rev} - \fr 8 9 T_{\rm cl}$.
In this case,
there are three nonzero terms in the expansion of the
wave function.
We find
\beq
\Psi (t_{\rm frac}) \approx
b_0 \psi_{\rm cl} (\fr 2 9)
+ b_1 \psi_{\rm cl} (\fr 8 9)
+ b_2 \psi_{\rm cl} (\fr 5 9)
\quad ,
\eeq
where the argument of $\ps_{\rm cl}$ is written
in units of $T_{\rm cl}$,
and where $\vert b_0 \vert \simeq 0.45$,
$\vert b_1 \vert \simeq 0.85$,
and $\vert b_2 \vert \simeq 0.29$.
This expansion predicts that the wave packet consists
of three subsidiary packets.
Since $\ps_{\rm cl}$
passes through the range 0 to $2\pi$ in $\ph$
at a constant rate,
at a specified fractional interval in the cycle
its angular position is at the corresponding
fractional part of $2 \pi$.
We therefore expect a large wave packet at
$\ph \simeq \fr 8 9 (2 \pi)$,
a medium one at $\ph \simeq \fr 2 9 (2 \pi)$,
and a small one at $\ph \simeq \fr 5 9 (2 \pi)$.
This prediction
agrees with the plot in Fig.\ 3e.
The wave packet at the time
$t_{\rm frac} + T_{\rm frac}$
is shown in Fig.\ 3f.
The periodicity $T_{\rm frac}$ of the wave packet is evident.

Since the wave function in this case has
three distinct subsidiary wave packets,
a large peak in the autocorrelation function is not to be expected.
This is compatible with the structure of Fig.\ 2b
at times near $t_{\rm frac} \simeq 27.6$ $\mu$sec.
The appearance of peaks of equal height
with a periodicity $\fr 1 2 T_{\rm frac}$
indicates that the fractional superrevival is evolving
into partial superrevivals twice each cycle,
which is possible because the subsidiary wave packets
are unequally weighted.

For $q=12$,
we find $l=12$, $N=8$, $\al = 79$,
$t_{\rm frac} = \fr 1 {12} t_{\rm sr} - \fr 3 {16} t_{\rm rev}$,
and $T_{\rm frac} = \fr 1 {4} t_{\rm rev} - \fr 5 {12} T_{\rm cl}$.
The expansion of the wave packet yields
\beq
\Psi (t_{\rm frac}) \approx
b_1 \psi_{\rm cl} (\fr 3 4)
+ b_4 \psi_{\rm cl} (\fr 1 2)
+ b_7 \psi_{\rm cl} (\fr 1 4)
+ b_{10} \psi_{\rm cl} (0)
\quad
\eeq
with $\vert b_1 \vert = \vert b_4 \vert
= \vert b_7 \vert = \vert b_{10} \vert \simeq \fr 1 2$.
The remaining $b_s$ coefficients vanish.
The packet in this case consists of
four equally weighted subsidiary packets evenly distributed
in $\ph$.
Figure 3g confirms this prediction.
However,
there is some distortion,
arising from higher-order terms
neglected in our analysis.
In Fig.\ 3h we plot the wave packet at time
$t_{\rm frac} + T_{\rm frac}$.
The periodicity is again evident.

For $q=15$,
we have $l=15$, $N=1$, $\al = 632$,
$t_{\rm frac} = \fr 1 {15} t_{\rm sr} - \fr 3 {20} t_{\rm rev}$,
and $T_{\rm frac} = \fr 1 {5}
t_{\rm rev} - \fr {11} {15} T_{\rm cl}$.
The expansion gives
\beq
\Psi (t_{\rm frac}) \approx
b_2 \psi_{\rm cl} (\fr 2 5)
+ b_5 \psi_{\rm cl} (\fr 4 5)
+ b_8 \psi_{\rm cl} (\fr 1 5)
+ b_{14} \psi_{\rm cl} (0)
\quad ,
\eeq
where the moduli of the complex coefficients are
$\vert b_2 \vert = \vert b_5 \vert \simeq 0.45$,
$\vert b_8 \vert \simeq 0.72$,
and $\vert b_{14} \vert \simeq 0.28$.
For these values,
we expect two equally weighted subsidiary wave packets
at $\fr 2 5 (2\pi )$ and $\fr 4 5 (2\pi )$,
a large subsidiary packet at $\fr 1 5 (2\pi )$,
and a small one at the origin.
Figure 3i displays the packet at $t_{\rm frac}$.
It agrees with the predictions.
Figure 3j shows the wave packet at the time
$t_{\rm frac} + T_{\rm frac}$,
which again exhibits the expected periodicity.

Finally,
in Fig.\ 3k,
we show the wave packet corresponding to $q=18$.
Here,
we obtain $l=6$, $N=8$, $\al = 79$,
$t_{\rm frac} = \fr 1 {18} t_{\rm sr} - \fr 1 8 t_{\rm rev}$,
and
$T_{\rm frac} = \fr 1 {6} t_{\rm rev} - \fr {17} {18} T_{\rm cl}$.
There are three nonzero coefficients $b_s$,
so
\beq
\Psi (t_{\rm frac}) \approx
b_0 \psi_{\rm cl} (\fr 1 9)
+ b_2 \psi_{\rm cl} (\fr 4 9)
+ b_4 \psi_{\rm cl} (\fr 7 9)
\quad ,
\eeq
with $\vert b_0 \vert \simeq 0.84$,
$\vert b_2 \vert \simeq 0.29$,
and $\vert b_4 \vert \simeq 0.45$.
A large subsidiary wave packet is visible near
$\fr 1 9(2\pi )$,
an intermediate-sized one is near $\fr 7 9(2\pi )$,
and a small one is near $\fr 4 9(2\pi )$,
in agreement with the theoretical result.
Figure 3l shows the wave packet at
$t_{\rm frac} + T_{\rm frac}$,
confirming that the periodicity is correctly
given by $T_{\rm frac}$.

For $q$ greater than $18$,
more than four subsidiary wave packets appear
in the expansion of the wave function
at the times $t_{\rm frac}$.
There are therefore fewer pronounced peaks
in the autocorrelation function,
and prominent partial superrevivals
are less likely to form.

\vglue 0.6cm
{\bf\noindent C. Radial Wave Packets}
\vglue 0.4cm

The previous sections have established the validity
of our theory for the various types of superrevival.
This subsection treats an example
closely related to a feasible experimental situation:
a p-state radial Rydberg wave packet for hydrogen
with ${\bar n} = 36$ and $\si = 1.5$.

Figure 4a shows the plot of the radial wave packet
at time $t=0$.
The wave packet starts its cycle at
the inner apsidal point and then moves
radially to the outer apsidal turning point.
Initially,
the wave packet is highly oscillatory.
As it moves,
the packet changes shape in a way characteristic
of a squeezed state
\cite{rss}.
In Fig.\ 4b,
we show the wave packet at the outer apsidal point
at the time $\fr 1 2 T_{\rm cl}$.
Here,
it is localized in the radial coordinate and is at
its point of minimum uncertainty.

At the full revival,
$t = t_{\rm rev}$,
the wave packet has reformed after undergoing its cycle
of collapse and fractional revivals and
is located at its initial position near the inner apsidal point.
For ease of comparison of the initial and full-revival packets,
we show in Fig.\ 4c the full-revival packet
displaced forward in time
by $\fr 1 2T_{\rm cl}$.
This brings it to the outer apsidal point
where it is the most localized.
The initial and full-revival packets are similar,
although the latter has
some distortion due to the higher-order corrections.

As before,
the full superrevival is predicted to occur for $q=6$.
This gives $l=6$, $N=72$, $\al = 1$,
$t_{\rm frac} = \fr 1 6 t_{\rm sr}$,
and $T_{\rm frac} = \fr 1 2 t_{\rm rev} - \fr 1 2 T_{\rm cl}$.
Note that since ${\bar n} = 36$ is divisible by four,
the fraction $m/n$ in $t_{\rm frac}$ vanishes.
The expansion of the wave function
is $\Psi (t_{\rm frac}) \approx b_4 \psi_{\rm cl} (\fr 2 3)$,
with $b_4 = 1$ and all the other $b_s$ coefficients vanishing.
In this case,
$\psi_{\rm cl}$ is a radial wave function,
which in one complete cycle moves from the inner
turning point to the outer one and back.
At $t_{\rm frac}$,
the wave function therefore represents a single packet
two-thirds through the classical cycle.
To compare this packet to the full revival,
we view it displaced backward
by $\fr 1 6T_{\rm cl}$ so it is at the outer turning point.
Figure 4d shows the wave packet at this time.
The full superrevival is visibly better
localized than the full revival
in Fig.\ 4c,
and it has less distortion.
In Fig.\ 4e,
the wave packet is displayed one period $T_{\rm frac}$ later
than the time of Fig.\ 4d.
Comparison shows that the packet has the
predicted period $T_{\rm frac}$.

\vglue 0.6cm
{\bf\noindent V. RYDBERG WAVE PACKETS IN ALKALI-METAL ATOMS}
\vglue 0.4cm

The previous sections of this paper have provided
a satisfactory theory of the long-term behavior
of Rydberg wave packets in hydrogen.
However,
the wave packets used to date in experiments
to exhibit full and fractional revivals
\cite{tenWolde,yeazell1,yeazell2,yeazell3,meacher}
have been produced in alkali-metal atoms.
In the remainder of this paper,
we show how to extend our previous theoretical
analysis to these cases.

The Rydberg series for an alkali-metal atom is given
by the energies
\beq
E_{n^\ast} = -\fr {1} {2 {n^\ast}^2}
\quad ,
\label{rydberg}
\eeq
where $n^\ast = n - \de (l)$,
and $\de (l)$ is an asymptotic quantum defect
for an alkali-Rydberg atom.
In what follows,
we consider Rydberg wave packets
in alkali-metal atoms
that can be represented as
a superposition of states strongly
peaked around a central value
${\bar n}^\ast = {\bar n} - \de (l)$
of the shifted principal quantum number.
Following the simplification introduced
at the end of Sec.\ IIA,
we assume $\bar n$ is an integer.
The extension of the results below to noninteger $\bar n$,
along with the issue of disentangling laser detuning
from features induced by quantum defects,
has been considered in
ref.\ \cite{detunings}.

We proceed by expanding the Rydberg energies as a
Taylor series in $n^\ast$,
\beq
E_{n^\ast} \simeq E_{{\bar n}^\ast}
+ E_{{\bar n}^\ast}^\prime (n - {\bar n})
+ \fr 1 2 E_{{\bar n}^\ast}^{\prime\prime} (n - {\bar n})^2
+ \fr 1 6 E_{{\bar n}^\ast}^{\prime\prime\prime} (n - {\bar n})^3
+ \cdots
\quad .
\label{energystar}
\eeq
The derivatives are taken with respect to $n^\ast$,
and we have used the relation
$(n^\ast - {\bar n}^\ast ) = (n - {\bar n})$.
The first three derivative terms define the time scales
\beq
T_{\rm cl}^{\ast} = 2 \pi  {\bar n}^{\ast 3}
\eeq
\beq
t_{\rm rev}^{\ast} = \fr {2 {\bar n}^{\ast}} 3 T_{\rm cl}^{\ast}
\eeq
and
\beq
t_{\rm sr}^{\ast} = \fr {3 {\bar n}^{\ast}} 4 t_{\rm rev}^{\ast}
\quad ,
\eeq
which depend on the quantum defects.

Since the quantum defects $\de (l)$ depend on $l$,
we may no longer treat circular and radial wave packets generically.
Circular wave packets have large values of $l$
and so vanishing values of $\de (l)$.
A hydrogenic description therefore
should be a good approximation for
the analysis of the behavior of
circular wave packets for alkali-metal atoms.
In contrast,
radial wave packets have small values of $l$.
This means their behavior
is modified by the presence of quantum defects,
with time scales and periodicities
changed relative to those in hydrogen.
In the following,
we limit ourselves to considering radial wave packets,
for which the quantum defects are important.

To incorporate non-hydrogenic features in our treatment,
we use supersymmetry-based quantum-defect theory (SQDT),
which has analytical wave functions with exact eigenvalues
reproducing the Rydberg series for alkali-metal atoms
\cite{sqdt}.
This model describes the behavior of the valence
electron in an alkali-metal atom as that of a single
particle in an effective central potential.
The effective potential is found by acting on the
radial Coulomb potential for hydrogen with a supersymmetry
transformation and then adding specific
supersymmetry-breaking terms
that incorporate electron-electron interactions
and reproduce the Rydberg series while
leaving the eigenfunctions analytical.
A modified angular quantum number
$l^{\ast} = l - \delta(l) + I(l)$
is introduced,
where $I(l)$ is an integer playing the role of the
supersymmetric shift.
The resulting SQDT hamiltonian is equivalent
to one obtained by replacing
$n,l,E_n$ in the radial equation by
$n^{\ast},l^{\ast},E_{n^{\ast}}$,
respectively.
The exact three-dimensional SQDT wave functions are
$Y_{l m}(\th,\ph) R_{{n^{\ast}} {l^{\ast}}}(r)$.
These have the eigenenergies $E_{n^\ast}$ given in
Eq.\ \rf{energystar}
and form a complete and orthonormal set.
More details on this model and references to the
recent literature may be found in the review article
\cite{susyreview}.

Let us denote by $\varphi_{\ast n} ({\vec r})$
the eigenstates of an alkali-metal atom
that are appropriate
for the description of a radial wave packet.
The analogue of Eq.\ \rf{wave1} for the wave function
of the packet is then given by
\beq
\Psi ({\vec r},t) = \sum_{n=1}^{\infty} c_{n^\ast}
\varphi_{\ast n} ({\vec r}) \exp \left[ -i E_{n^\ast} t \right]
\quad ,
\label{wavesqdt}
\eeq
where the $c_{n^\ast}$ are complex weights
determined by the initial wave function.
One role of SQDT in the analysis that follows
is to provide explicit analytical expressions
for the eigenstates in Eq.\ \rf{wavesqdt},
i.e., we take
$\varphi_{\ast n} ({\vec r})=
Y_{l m}(\th,\ph) R_{{n^{\ast}} {l^\ast}}(r) $.

Using the Taylor series \rf{energystar},
the wave function becomes
\beq
\Psi ({\vec r},t) = \sum_{k=-\infty}^{\infty} c_k
\varphi_{\ast k} ({\vec r}) \exp \left[ -2 \pi i
\left( \fr {kt} {T_{\rm cl}^{\ast}}
-\fr {k^2 t} {t_{\rm rev}^{\ast}}
+ \fr {k^3 t} {t_{\rm sr}^{\ast}} \right) \right]
\quad .
\label{bigpsistar}
\eeq
We have kept the first three terms in the Taylor expansion
so that times of order $t_{\rm sr}^{\ast}$ can be considered.
The sum ranges over the integer values of
$k = (n^{\ast} - {\bar n}^{\ast}) = (n - {\bar n})$.
We assume that ${\bar n}^\ast$ is large so that
the lower limit in the sum may be written as $- \infty$.
Since $k$ is integer valued,
we can take the coefficients $c_k$
as the gaussian functions defined for hydrogen in Eq.\ \rf{cn}.

We parametrize the times at which
subsidiary wave packets form as
\beq
t_{\rm frac}^{\ast} = \fr p q t_{\rm sr}^{\ast}
 - \fr m n t_{\rm rev}^{\ast}
\quad ,
\label{tfracstar}
\eeq
where $p$ and $q$ are relatively prime integers,
as are $m$ and $n$.
\it A priori, \rm
the integers $p$, $q$, $m$, $n$
are unconstrained,
and in particular they are \it not \rm
related to the corresponding
quantities for a hydrogenic radial wave packet.

Following the hydrogenic notation as closely as possible,
we express ${\bar n}^\ast$ as
\beq
{\bar n}^\ast = 4 \et + \la - \fr {\mu} {\nu}
\quad ,
\label{nbarstar}
\eeq
where $\et$, $\la$, $\mu$, $\nu$ are all integer valued,
$\la = 0$, $1$, $2$, or $3$,
and $\mu/\nu$ is an irreducible fraction
less than one.
In the most general case,
$\mu / \nu$ would represent the fractional part
of a combination of the quantum defect and a laser detuning.
In this paper,
we assume the laser is on resonance,
so $\mu / \nu$ is the fractional part of the quantum defect.
The effects of an additional laser detuning are
described in
ref.\ \cite{detunings}.
Note that writing the quantum defect as an
integer part minus a rational fraction of unity
can always be done
for experimentally determined quantum defects,
which of necessity have only a finite accuracy.

We also define a wave function
with time dependence involving only the
linear term in $k$.
In terms of $t_{\rm rev}^\ast$,
it is
\beq
\psi_{\rm cl} ({\vec r},t) = \sum_{k=-\infty}^{\infty} c_k
\varphi_{\ast k} ({\vec r}) \exp \left[ -2 \pi i
\fr {2 {\bar n}^\ast} 3 \fr {kt} {t_{\rm rev}^\ast} \right]
\quad .
\label{psiclstar}
\eeq
Note that at time zero,
$\Psi ({\vec r},0) = \psi_{\rm cl} ({\vec r},0)$.

In Appendic C,
we show that
the phases induced by the shifts $\De t$ in
$\psi_{\rm cl} ({\vec r},t)$
have the same period as the higher-order
contributions to the time-dependent phase in
$\Psi ({\vec r},t)$ at the times $t_{\rm frac}^\ast$.
It is therefore plausible to use the set
$\psi_{\rm cl} ({\vec r},t_{\rm frac}^\ast + \fr {s \al} l
T_{\rm cl}^\ast)$
with $s = 0, 1, \ldots, l-1$
as a basis for an expansion of $\Psi ({\vec r},t_{\rm frac}^\ast)$.
We write
\beq
\Psi ({\vec r},t_{\rm frac}^\ast) = \sum_{s=0}^{l-1} b_s
\psi_{\rm cl} ({\vec r},t_{\rm frac}^\ast
+ \fr {s \al} l T_{\rm cl}^\ast )
\quad .
\label{expansstar}
\eeq
It is shown in Appendix C that the $b_s$ coefficients
have the same form as in Eq.\ \rf{bs} for hydrogen.
The proof that this expansion is valid then follows as before,
using Eq.\ \rf{sum}.

Equation \rf{expansstar}
implies that the expansion
in terms of subsidiary packets
of a wave packet in an alkali-metal atom
is similar to a corresponding one in hydrogen.
In particular,
the number of subsidiary wave packets that form
at times $t_{\rm frac}^\ast$
and their relative proportions
are the same as those in a corresponding
wave packet in hydrogen at times $t_{\rm frac}$.
However,
the time scales
$t_{\rm sr}^\ast$, $t_{\rm rev}^\ast$, and $T_{\rm cl}^\ast$
controlling the behavior are different
from those in hydrogen,
as are the times $t_{\rm frac}^\ast$
at which the expansion in subsidiary packets is valid.

In Appendix C,
we determine the allowed values of $t_{\rm frac}^\ast$.
We find $q$ must again be restricted to multiples of 3.
The ratio $m/n$ is given in
Eq.\ \rf{mnstar}.

The proof that the wave packet and absolute square of
the autocorrelation function are periodic for times near
$t_{\rm frac}^\ast$,
with $p=1$,
is also outlined in Appendix C.
The allowed values of the period $T_{\rm frac}^\ast$
are found to be
\beq
T_{\rm frac}^\ast = \fr 3 q t_{\rm rev}^\ast
- \fr u v T_{\rm cl}^\ast
\quad ,
\label{Tfracstar}
\eeq
where the integers $u$ and $v$
are given in Eq.\ \rf{uvstar}.

Comparison of $u/v$ for alkali-metal atoms
(Eq.\ \rf{uvstar} of Appendix C)
with the corresponding definition for hydrogen
(Eq.\ \rf{uv2} of Appendix B),
reveals the appearance of an additional shift that
depends on the fractional part ${\mu}/{\nu}$
of ${\bar n}^\ast$.
This means that
the quantum defects cause an extra shift in the period
$T_{\rm frac}^\ast$,
in addition to the rescaling of
$t_{\rm rev}$ and $T_{\rm cl}$.
Note that if $\mu = 0$,
so that ${\bar n}^\ast$ is an integer,
then the analysis reduces to the hydrogenic case.

To summarize,
we have shown that a radial Rydberg wave packet
in an alkali-metal atom
forms superrevivals at times
$t_{\rm frac}^{\ast}$
given by Eq.\ \rf{tfracstar},
where $p=1$, $q$ is a multiple of three,
and the ratio $m/n$ is determined.
At these times,
the wave function gathers into a finite
number of subsidiary wave packets that move
periodically with period
$T_{\rm frac}^\ast$
given by Eq.\ \rf{Tfracstar}
with a specified ratio $u/v$.
The expressions for $t_{\rm frac}^\ast$ and $T_{\rm frac}^\ast$
differ from those for a corresponding
wave packet in hydrogen
by rescalings and shifts
that depend on the quantum defect.

\vglue 0.6cm
{\bf\noindent VI. EXAMPLES FOR RUBIDIUM}
\vglue 0.4cm

To illustrate in an experimentally viable
scenario the long-term behavior of Rydberg wave packets
when quantum defects are present,
we consider in this section an example
of a radial wave packet in rubidium
with ${\bar n} = 36$.
For definiteness,
we take p-state angular distributions,
corresponding to the wave packets produced
by a single short laser pulse.
The example illustrates that full and fractional
superrevivals may be observed for values of $\bar n$
corresponding to time delays currently
accessible in experiments.
Some other experimental issues are considered in the
next section.

In addition to providing an additional
check on the predictions of our theory,
the long-term revival structure
arising in this example
can be compared to the analogous situation for hydrogen
presented in Sec.\ IVC.
For the autocorrelation function
the behavior is generated
from Eq.\ \rf{auto}
with $E_n$ replaced with $E_{n^\ast}$.
For the radial wave packet itself,
it is obtained
from the analytical expression
\rf{wavesqdt}.
We use a gaussian distribution of width $\si = 1.5$
for the weighting coefficients $c_n$,
which makes it a good approximation
to truncate the sum over the SQDT eigenstates
after a finite number of terms.

The quantum defect for p states of rubidium
is $\de (1) \simeq 2.65$.
This gives ${\bar n}^\ast \simeq 33.35$,
$\eta = 8$, $\la = 2$,
and $\mu/\nu = 13/20$.
The time scales determining the behavior of this packet
are therefore
$T_{\rm cl}^\ast \simeq 5.63$ psec,
$t_{\rm rev}^\ast \simeq 0.12$ nsec,
and $t_{\rm sr}^\ast \simeq 3.13$ nsec.

Figure 5 shows the absolute square of the autocorrelation
function for this wave packet on time scales of order
$\fr 1 6 t_{\rm frac}^\ast$.
The full revival peak is visible near
$t_{\rm rev}^\ast$.
For the relatively small value of $\bar n$ considered here,
the revival structure is not as prominent as for larger
values.
Full and fractional superrevivals should nonetheless occur,
albeit with less visible features.
For $q=6$,
our theoretical analysis
predicts the occurrence of a full superrevival
at $t_{\rm frac}^\ast \simeq 0.50$ nsec
with periodicity $T_{\rm frac}^\ast \simeq 0.06$ nsec.
Figure 5 contains peaks
near $t_{\rm frac}^\ast$ in agreement with these values.
The theory also predicts the appearance of
fractional superrevivals
for $q=9$ and $12$ at the times
$t_{\rm frac}^\ast \simeq 0.33$ nsec and $0.25$ nsec,
respectively.
The corresponding predicted periodicities are
$T_{\rm frac}^\ast \simeq 0.04$ nsec and $0.03$ nsec.
Peaks with appropriate periodicities may be observed
near these times $t_{\rm frac}^\ast$ in Fig.\ 5.

Consider next the radial wave packet itself
for this example.
Fig.\ 6 displays the packet at different times.
The results should also be compared with
the figures for the corresponding hydrogenic packet
shown in Fig.\ 4.

The initial packet at time $t=0$
is displayed in Fig.\ 6a.
It is highly oscillatory and located
near the inner apsidal point.
Halfway through the classical orbit,
at time $t = \fr 1 2 T_{\rm cl}^\ast$,
the wave function is radially localized
and at its point of minimum uncertainty.
See Fig.\ 6b.
The wave packet near the full revival time
$t_{\rm rev}$ is shown in Fig.\ 6c,
after a time displacement by
$\fr 4 {15} T_{\rm cl}^\ast$.
This displacement brings the packet to the outer
apsidal point for ease of comparison.

The theoretical analysis
suggests that a full superrevival occurs at
$t_{\rm frac}^\ast = \fr 1 {6} t_{\rm sr}^\ast
- \fr {27} {160} t_{\rm rev}^\ast$
with period
$T_{\rm frac}^\ast = \fr 1 {2} t_{\rm rev}^\ast
- \fr {37} {60} T_{\rm cl}^\ast$.
The definition \rf{bs}
of the $b_s$ coefficients gives
$b_1 = 1$, while the other $b_s$ vanish.
Evaluating the expansion \rf{expansstar} yields
$\Psi (t_{\rm frac}) \approx
\psi_{\rm cl} (\fr 1 {10})$,
where the argument of $\ps_{\rm cl}$
is written in units of
$T_{\rm cl}^\ast$.
For purposes of comparison at the outer apsidal point,
we displace the packet by an additional amount
$\fr 2 5 T_{\rm cl}^\ast$.
The result is shown in Fig.\ 6d.
A single wave packet appears,
with less distortion than the full revival in Fig.\ 6c.

In Fig.\ 6e,
the wave packet is displayed a time $T_{\rm frac}^\ast$
later than in Fig.\ 6d.
The wave packet is again at the outer
apsidal point.
The motion has
period in agreement with $T_{\rm frac}^\ast$.

\vglue 0.6cm
{\bf\noindent VII. DISCUSSION}
\vglue 0.4cm

In this paper,
the behavior of Rydberg wave packets has been discussed
on time scales large compared
to the full-revival time $t_{\rm rev}$.
We have demonstrated the appearance
of various types of superrevival and have
presented a theoretical analysis that
correctly predicts the times of formation,
the associated periodicities,
and their structure.
Our analysis covers both hydrogenic
Rydberg wave packets and the experimentally
preferred Rydberg packets
in alkali-metal atoms.

The fractional superrevivals that we present
have long-term periodicities different
from those of the usual fractional revivals.
These periodicities could be observed for values of $\bar n$
corresponding to time delays in a pump-probe experiment
of under one nanosecond.

When several packets form,
we have shown that
they often evolve quickly into a state
where one packet is much larger than the others.
In this case,
large peaks appear in the autocorrelation function.
All these features have been confirmed
in numerically computed
examples both for the autocorrelation function
and for the wave packets themselves.

The wave-packet behavior on the time scale
$t_{\rm sr}$
is self-similar in some ways to
the fractional-revival structure
appearing on the time scale $t_{\rm rev}$.
There are significant and experimentally
observable differences,
however.
For example,
the times $t_{\rm frac}$ are more restricted than the
times at which the usual fractional revivals occur,
since only times approximately given by
certain irreducible fractions
of $\fr 1 3t_{\rm sr}$ are permitted.
Moreover,
the allowed periodicities $T_{\rm frac}$ are
dominated by multiples of $3 t_{\rm rev}$.
The origin of the frequent appearance of multiples of three
and one-third in our theory can be traced
to the appearance of the third-order derivative of the energy
in the original wave-packet expansion.

Our generalization of the hydrogenic analysis
to include quantum defects uses
a supersymmetry-based quantum-defect theory
that provides a complete and orthonormal
set of analytical wave functions
with exact Rydberg eigenenergies.
We have thereby proved that
an expansion in terms of subsidiary
packets is valid for a radial Rydberg packet
in an alkali-metal atom.
The controlling time scales $T_{\rm cl}$, $t_{\rm rev}$, and
$t_{\rm sr}$ are modified relative to the
hydrogenic case by quantum defects,
as are the periodicities $T_{\rm frac}$ of the motion
and the times $t_{\rm frac}$
at which the subsidiary-packet expansion is valid.
Since the quantum defects for p states of the heavier
alkali-metal atoms can be relatively large,
these modifications can be significant
and must be included in any accurate
description of the evolution of the wave packet.

Relatively little prior literature exists
on the long-term behavior of Rydberg wave packets.
A long-term revival at
$t=\bar n^3 T_{\rm cl}$
has been presented in ref.\ \cite{ps}.
The time-dependent autocorrelation function \rf{auto}
is an example of an almost-periodic function if
the sum is truncated to a finite number of terms.
Discussions of such functions exist in
a more general context
(see, for example, ref.\ \cite{quasi}).
However,
the wave function expansion \rf{wavesqdt}
has spatial dependence that is critical in
the superrevival structure,
involving the formation of the subsidiary wave packets.
Ref.\ \cite{peres} discusses
a class of long-term revivals in hydrogen
occurring when $t$ and $\bar n$ are such that
the energy expansion yields a phase that is an
integer multiple of $2\pi i$.
In contrast,
the analysis of the full/fractional superrevivals
presented here,
which holds for arbitrary values of $\bar n$
in any alkali-metal atom,
is based on finding the times
$t_{\rm frac}$ where the phase in $\Psi ({\vec r},t)$
matches the phases of the shifted waves $\psi_{\rm cl}$
in the expansion.
We find that at $t_{\rm frac}$ for $q=6$,
corresponding to the full superrevival,
the phase of the time-dependent terms
is not necessarily
an integer multiple of $2 \pi i$.
The point is that
the superrevival times we find
correspond to the formation of subsidiary wave packets.
Unlike the usual fractional revivals,
these are \it not \rm equally weighted.
The counterintuitive appearance of the
$q=6$ superrevival occurs because all but
one of the weighting coefficients $b_s$ vanish.

In the remainder of this section,
we discuss some issues
relevant to the experimental
verification of these results.
Consider a pump-probe measurement utilizing
either time-delayed photoionization
or phase modulation
with single-photon excitation,
yielding a wave packet with p-state angular distribution.
In time-delayed photoionization,
the packet evolves after formation
for a delay time $t$,
whereupon it interacts with a probe pulse ionizing the atom.
In the phase-modulation technique,
the wave packet is excited by two identical laser pulses
separated in time by a delay $t$.
The interference of the two wave packets produces
an ionization signal proportional to their overlap.

Both these approaches permit measurement of
an ionization signal displaying periodicities
corresponding to the full/fractional revivals.
Refs.\ \cite{yeazell2,yeazell3} describe
the observation of a
time-delayed photoionization signal for
p-state radial wave packets in potassium
for delay times up to approximately 800 psec.
The measured ionization signal reveals both
full and fractional revivals without any
appreciable loss of the signal.
Ref.\ \cite{wals} uses the phase-modulation technique
for wave packets of rubidium with ${\bar n} \simeq 46.5$
and $53.3$.
Fractional revivals with periodicities as small
as $\fr 1 7 T_{\rm cl}$ have been resolved.
However,
there was a decay of the signal envelope with
increasing delay,
and the signal was quite small after 300 psec.

The limiting factor for the time delay in these
experiments appears to be the ability to
maintain a good overlap of the two laser pulses in
the interaction region as the delay is increased.
For Rydberg states with $\bar n = 25$ -- $50$,
transitions induced by interactions with black-body
radiation are not an issue in this context
\cite{black-body,gallagher}.
Although the Rydberg-state lifetimes
induced by black-body radiation
are typically reduced by more
than an order of magnitude compared to
lifetimes induced by spontaneous emission,
they are still several orders of
magnitude more than
the superrevival time scale $t_{\rm sr}$.
With a sufficiently low background pressure,
transitions induced by collisional processes should
not be a factor either.

The time-delayed photoionization experiments
in refs.\ \cite{yeazell2,yeazell3}
maintained a good signal for time
delays of up to 800 psec.
This is long enough to allow detection
of both full and fractional superrevivals.
In the example of rubidium
with ${\bar n} = 36$,
considered in Sec.\ VI,
$t_{\rm frac}^\ast$ for $q=6$ was found to be
approximately 500 psec.
Therefore,
this full superrevival should be observable with
existing experimental capabilities.
The fractional superrevival at
$t_{\rm frac}^\ast \simeq 250$ psec with
$T_{\rm frac}^\ast \approx \fr 1 4 t_{\rm rev}$
should be measurable as well.

As $\bar n$ is increased,
the number of autocorrelation
peaks increases and the fractional
superrevivals become more prominent.
Values of ${\bar n} \simeq 50$ would require
a delay line of 3 -- 4 nsec,
and the overlap of the laser pulses would have
to be maintained for this amount of time.
Assuming this can be achieved,
the observation of the superrevival in this
case would show the return to the near classical
state of a quantum wave packet after a time
of over twelve hundred classical orbits.

\vglue 0.6cm
{\bf\noindent ACKNOWLEDGMENTS}
\vglue 0.4cm

R.B. thanks Colby College for a Science Division Grant.
V.A.K. thanks the Aspen Center for Physics
for hospitality during the initial stages of this work.

\vglue 0.6cm
{\bf\noindent APPENDIX A}
\vglue 0.4cm

In this appendix,
we show that at the times $t_{\rm frac}$
given in Eq.\ \rf{times}
the wave function $\Psi ({\vec r},t)$ can be expanded
as a sum of subsidiary waves of the form
$\psi_{\rm cl} ({\vec r},t)$,
with $t$ shifted by a fraction of the
period $T_{\rm cl}$.
Section A1 analyzes the phase periodicity of
$\Psi ({\vec r},t)$ at the times $t_{\rm frac}$,
while Sec.\ A2 obtains the phase periodicities
of the subsidiary wave functions in the sum
and matches them with the periodicity of
$\Psi ({\vec r},t)$.
We then examine the constraints on the times
$t_{\rm frac}$ in A3 and the properties of the expansion
coefficients in A4.

\vglue 0.6cm
{\bf\noindent 1. Phase of Higher-Order Contributions
in $\Psi ({\vec r},t)$}
\vglue 0.4cm

We first examine the periodicity
at the times $t_{\rm frac}$
of the phase of the full wave function
$\Psi ({\vec r},t)$.
We are interested in the terms of higher order
in $k$ in Eq.\ \rf{psi3rd}
that result from the expansion of the
energy in a Taylor series.

Substituting the expression for $t_{\rm frac}$ in
Eq.\ \rf{times}
into $\Psi ({\vec r},t)$
and neglecting terms in the exponential of order
$t_{\rm rev}/t_{\rm sr} \ll 1$,
we find that the second and third-order terms in
the energy lead to a phase factor
$\exp \left[ 2 \pi i \th_k \right]$,
where
\beq
\th_k = \fr {3 {\bar n} p} {4 q} k^2
- \fr m n k^2 - \fr p q k^3
\quad .
\label{thetak}
\eeq
We wish to find the minimum
period $l$ for shifts in $k$ that leaves
the phase $\exp \left[ 2 \pi i \th_k \right]$ invariant.

The analysis is simplified if we first impose a condition
on $m/n$ to reduce the general case to the same form
as the one with $\bar n$ divisible by four.
The definition for $\bar n$ in
Eq.\ \rf{nbar}
suggests we impose
\beq
\fr m n =
\fr {3 \la p} {4 q}
\quad\quad {\rm (mod \, 1)}
\quad ,
\label{mn}
\eeq
which simplifies the expression for $\th_k$ in the general case.
The relatively prime integers $m$ and $n$
obeying Eq.\ \rf{mn}
are then specified once the right-hand side of Eq.\ \rf{mn}
has been fully reduced.

The choice \rf{mn}
leads to a simplified form of the phase:
\beq
\th_k = \fr {3 \et p} {q} k^2
- \fr p q k^3
\quad .
\label{etatheta}
\eeq
Define $l$ as the minimum period in $k$ that leaves
$\th_k$ invariant (mod 1),
i.e., $l$ is the smallest integer
satisfying $\th_{k+l} = \th_k$ (mod 1).
Then,
there are two possible solutions for $l$,
both dependent on $q$:
\beq
l =\cases{q&if~~$q/9 \ne 0~~ ({\rm mod}~1)~~$,\cr
        q/3&if~~$q/9 = 0~~ ({\rm mod}~1)~~$.\cr }
\quad
\label{lvalues}
\eeq

\vglue 0.6cm
{\bf\noindent 2. Phase of the Subsidiary Waves
$\psi_{\rm cl} ({\vec r},t)$}
\vglue 0.4cm

The second phase of interest results when the macroscopically
distinct subsidiary waves $\ps_{\rm cl}$ are shifted in time.
We next evaluate the periodicity of this phase and
match it to the one obtained above for $\Psi ({\vec r},t)$.

Consider a shift $\De t$ in time of the classical
wave function $\psi_{\rm cl} ({\vec r},t)$ defined in
Eq.\ \rf{psicl}.
We take a shift of the form
\beq
\De t = \fr a b t_{\rm rev} + \fr c d T_{\rm cl}
\quad ,
\label{timeshift}
\eeq
where $a$ and $b$ are relatively prime integers,
as are $c$ and $d$.
This shift generates a phase factor
$\exp \left[ - 2 \pi i \ph_k \right]$
in $\psi_{\rm cl} ({\vec r},t)$,
where
\beq
\ph_k = \fr {2 {\bar n} a} {3 b} k
+ \fr c d k
\quad .
\label{phi1}
\eeq

To simplify this expression,
we choose the relatively prime integers $c$ and $d$
to satisfy
\beq
\fr {2 \la a} {3 b}
+ \fr c d = 0
\quad\quad {\rm (mod \, 1)}
\quad ,
\label{cd}
\eeq
where the first term is understood to be fully reduced.
The phase $\ph_k$ then reduces to
\beq
\ph_k = \fr {8 \et a} {3 b} k
\quad .
\label{phi2}
\eeq

To write the full wave function as an expansion
in the subsidiary waves,
the two sets of phases must have the same periodicity.
We therefore impose the condition that the
phase $\ph_k$ be periodic in $k$ with the
same minimum value $l$ as the period of the
phase $\th_k$ in $\Psi ({\vec r},t)$.
This requirement can be satisfied
by an appropriate reduction of the fractional coefficient
of $k$ in Eq.\ \rf{phi2}.
It implies that we can write
\beq
\fr {8 \et a} {3 b} = \fr {\al N} {l N}
= \fr {\al} l
\quad .
\label{reduce}
\eeq
The integer $N$ is the product of all factors of $8 \et a$
that are also factors of $l$,
and $\al = {8 \et a}/N$ is also an integer.
Since $\al$ contains no factors that are also factors of $l$,
$\al$ and $l$ are relatively prime.
With this reduction,
$\ph_k = \al k/l$.
The minimum period for $k$ in $\ph_k$ is therefore $l$,
as desired.

With the choices for $c$ and $d$ in
Eq. \rf{cd}
and the reduction of ${8 \et a}/{3 b}$ in
Eq.\ \rf{reduce},
we find that the time shifts $\De t$ in
$\psi_{\rm cl} ({\vec r},t)$ become
\beq
\De t = \fr {\al} l T_{\rm cl}
\quad .
\label{deltat}
\eeq

\vglue 0.6cm
{\bf\noindent 3. Constraints on $t_{\rm frac}$}
\vglue 0.4cm

The expansion of the wave function $\Psi ({\vec r},t)$
in terms of the shifted wave functions $\psi_{\rm cl}$
is given in
Eq.\ \rf{expans},
and the $b_s$ coefficients are defined in
Eq.\ \rf{bs}.
The time of formation of the subsidiary wave
packets are the times
$t_{\rm frac}$
given by Eq.\ \rf{times}.
The integers $m$ and $n$ must satisfy
Eq.\ \rf{mn}.
We next prove that $l$ and hence $q$ must be
a multiple of three.

Examining the reduction of the ratio in
Eq.\ \rf{reduce},
we see that the integer $b$ satisfies the condition
$b = lN/3$.
Since all the factors of $N$ are also factors of $l$,
the only way that $b$ can be an integer is if $l$
(and hence $N$ as well) is a multiple of three.
Therefore,
the allowed values of $l$ must all be multiples of three.
Since $l$ is given in Eq.\ \rf{lvalues} either as $q$ for
$q/9 \ne 0$ (mod 1) or as $q/3$ for
$q/9 = 0$ (mod 1),
it follows that $q$ must also
be a multiple of three.

It turns out
that the analysis of the wave-packet
and autocorrelation-function periodicities given in
Appendix B
yields the further constraint $p=1$.
We therefore have the constraints on the times
$t_{\rm frac}$ given by Eq.\ \rf{times}
that $q$ must be a multiple of three,
$p=1$,
and $m/n$ obeys Eq.\ \rf{mn}.

\vglue 0.6cm
{\bf\noindent 4. Properties of $b_s$ Coefficients}
\vglue 0.4cm

The form of the
complex coefficients $b_s$ multiplying the subsidiary
wave functions $\psi_{\rm cl}$
at the times $t_{\rm frac}$,
given in
Eq.\ \rf{bs},
reduce the expansion \rf{expans}
to an identity.
Substitute Eq.\ \rf{bs}
into the expansion \rf{expans}
for $\Psi ({\vec r},t)$ at the times $t_{\rm frac}$,
and use the identity
\beq
\fr 1 l \sum_{s =0}^{l-1}
\Bigg( \exp \left[ -2 \pi i \fr {\al (k - k^\prime)} l
\right] \Bigg)^s
= \left( \de_{k,k^\prime} + \de_{k,k^\prime \pm l} + \ldots \right)
\quad .
\label{sum}
\eeq
This yields the expression in
Eq.\ \rf{psi3rd}
at times $t \approx t_{\rm frac}$,
up to terms in the exponential of order
$t_{\rm rev}/t_{\rm sr} \ll 1$.

Note that the identity \rf{sum}
holds only if $\al$ and $l$ are relatively prime.
This means that
the reduction of the ratio of integers in
Eq.\ \rf{reduce} is a necessary condition for the
formation of the subsidiary wave packets at the time
$t_{\rm frac}$.
Moreover,
as discussed in A3,
it leads to the restriction that $q$ must be a multiple of three.

If we multiply the expansion for $b_s$ in
Eq.\ \rf{bs}
by its complex conjugate,
sum over $s$,
and use the identity in
Eq.\ \rf{sum},
we get
\beq
\sum_{s =0}^{l-1} \vert b_s \vert^2 = 1
\quad .
\label{bssum}
\eeq
This implies that the normalization of the wave function
is maintained at the times $t_{\rm frac}$,
as expected.

\vglue 0.6cm
{\bf\noindent APPENDIX B}
\vglue 0.4cm

In this appendix,
we examine the periodicities of the wave packet
and the autocorrelation function.
Section B1 contains the
proof that the wave packet
$\vert \Psi ({\vec r},t) \vert^2$
is periodic at times $t \approx t_{\rm frac}$,
with a period $T_{\rm frac}$ that depends on
$t_{\rm rev}$ and $T_{\rm cl}$.
Section B2 contains the corresponding analysis
for the autocorrelation function.

\vglue 0.6cm
{\bf\noindent 1. Periodicity of the Wave Packet}
\vglue 0.4cm

The wave function $\Psi ({\vec r},t)$
through third order in the energy expansion
is given at the times $t_{\rm frac}$
by Eq.\ \rf{psi3rd}.
The issue of its periodicity can be addressed
by considering the effect of shifting the time,
\beq
t_{\rm frac} \rightarrow t_{\rm frac} + \fr e f t_{\rm rev}
\quad ,
\label{shift}
\eeq
where $e$ and $f$ are integers.

We are interested in the
smallest possible value of $e/f$
that leads to periodicity in
$\Psi ({\vec r},t)$.
The appearance of periodic structures in
$\Psi ({\vec r},t)$
upon time shifts of the form \rf{shift}
is possible only if the minimum periodicity
of $k$ in $\th_k$ (cf.\ Eq.\ \rf{etatheta})
remains $l$
\it and \rm
if the ensuing modified coefficients
in the expansion \rf{expans} can be expressed
in terms of the original coefficients $b_s$.
We have performed an analysis
for arbitrary $e$ and $f$
along the lines of what follows in this subsection,
from which we have shown that the minimum value of $e/f$ is
\beq
\fr e f = \fr 3 q
\quad .
\label{ef}
\eeq
For simplicity in what follows,
we restrict our treatment here to
this explicit value.

Under the shift given by Eqs.\ \rf{shift} and \rf{ef},
the phase $\th_k$ of the higher-order
time-dependent terms in $\Psi ({\vec r},t)$ becomes
\beq
\th_k^\prime = \fr {3 (\et p + 1)} {q} k^2
- \fr p q k^3
\quad ,
\label{thetaprime}
\eeq
where we have kept terms only up to order
${t_{\rm rev}}/{t_{\rm sr}} \ll 1$.
This phase continues to have the same minimum
period $l$ under changes in the summation index
$k$ as did the phase $\th_k$,
i.e.,
$\th_{k+l}^\prime = \th_k^\prime$ for the
same values of $l$ as in Eq.\ \rf{lvalues}.

Next,
consider the expansion of
$\Psi ({\vec r},t_{\rm frac} + \fr 3 q t_{\rm rev})$
as a sum of subsidiary wave functions
$\psi_{\rm cl}$
at the time $t_{\rm frac} + \fr 3 q t_{\rm rev}$:
\beq
\Psi ({\vec r},t_{\rm frac} + \fr 3 q t_{\rm rev})
 = \sum_{s=0}^{l-1} b_s^\prime
\psi_{\rm cl} ({\vec r},
t_{\rm frac} + \big( \fr {2 {\bar n}} q
+ \fr {s \al} l \big) T_{\rm cl} )
\quad .
\label{psi3trev}
\eeq
In this equation,
we have used the replacement $\fr 3 q t_{\rm rev}
= \fr {2 {\bar n}} q T_{\rm cl}$,
and the $b_s^\prime$ are complex coefficients given by
\beq
b_s^\prime = \fr 1 l \sum_{k^\prime =0}^{l-1}
\exp \left[ 2 \pi i \fr {\al s} l k^\prime \right]
\exp \left[ 2 \pi i \th_{k^\prime}^\prime \right]
\quad .
\label{bsprime}
\eeq
These coefficients are the same as
those of Eq.\ \rf{bs}
but with $\th_k \rightarrow \th_k^\prime$.

We next show that the set of coefficients $b_s^\prime$ in
Eq.\ \rf{bsprime}
can be found in terms of the previous set $b_s$,
which are defined in
Eq.\ \rf{bs}.
First,
we let $k^\prime \rightarrow k^\prime - 1$ in
the definition of $b_s^\prime$.
We get
$$b_s^\prime =
\exp \left[ 2 \pi i \fr {\al s} l \right]
\exp \left[ 2 \pi i \fr {3 (\et p + 1)} q \right]
\exp \left[- 2 \pi i \fr p q \right]
\, \fr 1 l \sum_{k = -1}^{l-2}
\exp \left[ 2 \pi i \fr {\al s} l k \right]
$$
\beq
\quad\quad\quad
\times  \exp \left[ 2 \pi i \left( \fr {3 (\et p + 1) - 3p} q k^2
- \fr p q k^3 \right) \right]
\exp \left[ 2 \pi i \left( \fr {6 (\et p + 1)} q
- \fr {3p} q \right) k \right]
\quad .
\label{bsprime2}
\eeq
If we require that $p = 1$,
then
\beq
\exp \left[ 2 \pi i \left( \fr {3 (\et p + 1) - 3p} q k^2
- \fr p q k^3 \right) \right]
= \exp \left[ 2 \pi i \th_k \right]
\quad ,
\label{pcond}
\eeq
where $\th_k$ is defined in
Eq.\ \rf{etatheta}.
We then define a quantity $J$ as:
\beq
J =\cases{3&if~~$q/9 \ne 0~~ ({\rm mod}~1)~~$,\cr
        1&if~~$q/9 = 0~~ ({\rm mod}~1)~~$,\cr }
\quad
\label{Jvalues}
\eeq
which permits us to write $q=3 l/J$.
This gives
$$b_s^\prime =
\exp \left[ 2 \pi i \fr {\al s} l \right]
\exp \left[ 2 \pi i \fr {3 (\et + 1)} q \right]
\exp \left[- 2 \pi i \fr 1 q \right]
\, \fr 1 l \sum_{k = -1}^{l-2}
\exp \left[ 2 \pi i \th_k \right]
$$
\beq
\quad\quad\quad\quad
\times
\exp \left[ 2 \pi i \left( \fr {\al s} l
+ \fr {J (2 \et + 1)} l \right) k \right]
\quad .
\label{bsprime3}
\eeq

The quantity $J (2 \et + 1)/l$ can be rewritten
as ${\al x}/l$ provided that an integer $x$
exists that satisfies
\beq
\al x = J (2 \et + 1)
\quad\quad {({\rm mod} \,\, l)}
\quad .
\label{xcond}
\eeq
It can be shown that an integer-valued equation of this
form has a solution $x$ if the greatest common divisor
of $\al$ and $l$ divides evenly into $J (2 \et + 1)$.
Since $\al$ and $l$ are relatively prime,
their greatest common divisor is 1,
and this divides evenly into $J (2 \et + 1)$.
Thus,
there always exists an integer $x$ that satisfies
Eq.\ \rf{xcond},
and we may replace $\left( {\al s}
+ {J (2 \et + 1)}\right) /l$ by
${\al (s+x)}/l$.

Since the exponential factors
inside the sum \rf{bsprime3} are periodic with period $l$,
we can shift the summation over $k$ back to the
range $k=0$ to $k=l-1$,
We then obtain
\beq
b_s^\prime =
\exp \left[ 2 \pi i \fr {\al s} l \right]
\exp \left[ 2 \pi i \fr {3 (\et + 1)} q \right]
\exp \left[- 2 \pi i \fr 1 q \right]
b_{(s+x)_{{\rm mod} \, l}}
\quad .
\label{bsprime4}
\eeq
Defining the phase factor
\beq
\exp \big[ 2 \pi i \Th_{s} \big] =
\exp \left[ 2 \pi i \fr {\al s} l \right]
\exp \left[ 2 \pi i \fr {3 (\et + 1)} q \right]
\exp \left[- 2 \pi i \fr 1 q \right]
\quad ,
\label{bigtheta}
\eeq
we then have for $p=1$ that
the coefficients $b_s^\prime$ are
given by
\beq
b_s^\prime =
\exp \big[ 2 \pi i \Th_s \big] \,
b_{(s+x)_{{\rm mod} \, l}}
\quad ,
\label{bsbigtheta}
\eeq
where $x$ is a given integer.
{}From this it follows that
$\vert b_s^\prime \vert = \vert b_{(s+x)_{{\rm mod} \, l}} \vert$.
We use this relation to prove that the motion
of the wave packet is periodic.

The probability density for the wave packet at the
time $t_{\rm frac} + \fr 3 q t_{\rm rev}$
is the absolute square of the expression \rf{psi3trev}.
Since the subsidiary wave packets are
by construction localized
in space and distributed evenly along the orbit,
they have little or no overlap.
This permits us to ignore the cross terms in the
absolute square of the wave function.
Substituting the expression \rf{bsbigtheta}
for $b_s^\prime$ into Eq.\ \rf{psi3trev}
and taking the absolute square gives
\beq
\vert \Psi ({\vec r},t_{\rm frac} + \fr 3 q t_{\rm rev}) \vert^2
 = \sum_{s=0}^{l-1} \vert b_{(s+x)_{{\rm mod} \, l}} \vert^2
\vert \psi_{\rm cl} ({\vec r},
t_{\rm frac} + \big( \fr {2 {\bar n}} q
+ \fr {s \al} l \big) T_{\rm cl} )
\vert^2
\quad .
\label{Psi1}
\eeq
Since the integer $x$ always exists,
we can set $s^\prime = (s+x)_{{\rm mod} \, l}$
and reorder the terms to get
\beq
\vert \Psi ({\vec r},t_{\rm frac} + \fr 3 q t_{\rm rev}) \vert^2
 = \sum_{s^\prime = 0}^{l-1} \vert b_{s^\prime} \vert^2
\vert \psi_{\rm cl} ({\vec r},
t_{\rm frac} + \big( \fr {2 {\bar n}}
q - \fr {\al x} l \big) T_{\rm cl}
+ \fr {\al s^\prime} l T_{\rm cl})
\vert^2
\quad .
\label{Psi2}
\eeq
Introduce the irreducible ratio $u/v$
of integers $u$ and $v$ satisfying
\beq
\fr u v = \fr {2 {\bar n}} q
- \fr {\al x} l
\quad\quad {\rm (mod \, 1)}
\quad .
\label{uv}
\eeq
Using
Eq.\ \rf{nbar}
of Sec.\ IIIA
and
Eqs.\ \rf{Jvalues} and \rf{xcond},
this definition of $u/v$ may be rewritten as
\beq
\fr u v = \fr {2 (\et + \la) - 3} q
\quad\quad {\rm (mod \, 1)}
\quad .
\label{uv2}
\eeq
Equation \rf{Psi2} then becomes
\beq
\vert \Psi ({\vec r},t_{\rm frac} + \fr 3 q t_{\rm rev}) \vert^2
=\vert \Psi ({\vec r},t_{\rm frac} + \fr u v T_{\rm cl}) \vert^2
\quad ,
\label{Psi3}
\eeq
where we have used
Eq.\ \rf{expans}
and the assumption that the subsidiary wave packets
do not overlap.

Since $T_{\rm cl} \ll t_{\rm rev}$,
the expansion in
Eq.\ \rf{expans}
holds for shifts in time of order $T_{\rm cl}$.
We may therefore subtract $\fr u v T_{\rm cl}$ from $t$
on both sides of
Eq.\ \rf{Psi3}
to obtain
\beq
\vert \Psi ({\vec r},t_{\rm frac} + \fr 3 q t_{\rm rev}
 - \fr u v T_{\rm cl}) \vert^2
 = \vert \Psi ({\vec r},t_{\rm frac}) \vert^2
\quad .
\label{periodicwave}
\eeq
Finally, we can define the period $T_{\rm frac}$
as in Eq.\ \rf{Tfrac},
where the integers $u$ and $v$ satisfy
Eq.\ \rf{uv2}.

We have thus proved that
at the times $t \approx t_{\rm frac}$,
the wave packet is periodic with the period
$T_{\rm frac}$.

\vglue 0.6cm
{\bf\noindent 2. Periodicity of the Autocorrelation Function}
\vglue 0.4cm

A similar method
can be used to show that at the times $t_{\rm frac}$
the absolute square of the autocorrelation function
is periodic with the same period
$T_{\rm frac}$.
The autocorrelation function at time $t_{\rm frac}$ is
$A(t_{\rm frac}) = \big< \Psi ({\vec r},0)
\vert \Psi ({\vec r},t_{\rm frac}) \big>$.
The initial wave function $\Psi ({\vec r},0)$ is
$\psi_{\rm cl} ({\vec r},0)$.
Shifting $t_{\rm frac}$ to
$t_{\rm frac} + \fr 3 q t_{\rm rev}$ and using
Eqs.\ \rf{psi3trev} and \rf{bsbigtheta}
and the definition of $u/v$ in
Eq.\ \rf{uv2} yields
\beq
A(t_{\rm frac} + \fr 3 q t_{\rm rev})
 = \sum_{s^\prime = 0}^{l-1}
\exp \big[ 2 \pi i \Th_{s^\prime} \big] \,
b_{s^\prime} \,
\big< \psi_{\rm cl} (0) \vert \psi_{\rm cl} (t_{\rm frac} +
\fr u v T_{\rm cl} + \fr {\al s^\prime} l T_{\rm cl}) \big>
\quad ,
\label{auto1}
\eeq
where we have suppressed the $\vec r$-dependence.
The phase
$\exp \big[ 2 \pi i \Th_{s^\prime} \big]$
is given in Sec.\ B1.
Note that the substitution $s^\prime = (s+x)_{mod \, l}$
has been made.

Taking the absolute square of
$A(t_{\rm frac} + \fr 3 q t_{\rm rev})$,
gives
$$
\vert A(t_{\rm frac} + \fr 3 q t_{\rm rev}) \vert^2
 = \sum_{s^\prime = 0}^{l-1}
\vert b_{s^\prime} \vert^2 \,
\vert \big< \psi_{\rm cl} (0) \vert \psi_{\rm cl} (t_{\rm frac} +
\fr u v T_{\rm cl} + \fr {\al s^\prime} l T_{\rm cl}) \big>
\vert^2
$$
$$
\quad
+ {\sum\sum}_{s^\prime \ne s = 0}^{l-1} \,
\exp \big[ 2 \pi i (\Th_{s^\prime} - \Th_{s}) \big] \,
b_{s^\prime} b_{s^\prime}^\ast \,
\big< \psi_{\rm cl} (t_{\rm frac} + \fr u v T_{\rm cl}
+ \fr {\al s} l T_{\rm cl}) \vert \psi_{\rm cl}(0) \big>
$$
\beq
\quad\quad\quad
\times
\big< \psi_{\rm cl} (0) \vert \psi_{\rm cl}
(t_{\rm frac} + \fr u v T_{\rm cl}
+ \fr {\al s^\prime} l T_{\rm cl}) \big>
\quad .
\label{bigauto}
\eeq
The first term on the right-hand side is just
$\vert A(t_{\rm frac} + \fr u v T_{\rm cl}) \vert^2$.
Since $s \ne s^\prime$ in the second term
and since we assume that the subsidiary wave functions do
not overlap appreciably,
one of the inner products in the double sum must vanish.
We can therefore neglect this term.

Subtracting $\fr u v T_{\rm cl}$ from both sides
produces
\beq
\vert A(t_{\rm frac} + \fr 3 q t_{\rm rev}
- \fr u v T_{\rm cl}) \vert^2
= \vert A(t_{\rm frac}) \vert^2
\quad .
\label{periodicauto}
\eeq
This proves that the square of the autocorrelation
function is periodic at the
times $t_{\rm frac}$ with period
$T_{\rm frac}$
given by Eq.\ \rf{Tfrac}.

\vglue 0.6cm
{\bf\noindent APPENDIX C}
\vglue 0.4cm

In this appendix,
we prove that at the times $t_{\rm frac}^\ast$
the SQDT wave packet $\Psi ({\vec r},t)$
for an alkali-metal atom
can be written as an expansion over
subsidiary packets $\psi_{\rm cl} ({\vec r},t)$.
In analogy with the method for hydrogen,
we proceed by matching the period of the phase
of the higher-order contributions in
$\Psi ({\vec r},t)$
with the period of the phase induced by shifting
$\psi_{\rm cl} ({\vec r},t)$ by a fraction of its
period $T_{\rm cl}^\ast$.
We then determine the coefficients in the expansion
and the constraints on the times $t_{\rm frac}^\ast$.
Lastly,
we show that the wave packet and autocorrelation
function are periodic with a period
$T_{\rm frac}^\ast$ and we determine the
allowed values of $T_{\rm frac}^\ast$.

Consider $\Psi ({\vec r},t)$ at the times
$t_{\rm frac}^\ast$.
An additional phase $\exp \left[ 2 \pi i \th_k^\ast \right]$
due to the higher-order corrections
appears relative to the linear term,
where
\beq
\th_k^\ast = \fr {3 {{\bar n}^\ast} p} {4 q} k^2
- \fr m n k^2 - \fr p q k^3
\quad .
\label{thetakstar}
\eeq
As usual,
terms of order
${t_{\rm rev}^\ast}/{t_{\rm sr}^\ast} \ll 1$
in the exponential have been neglected.

To simplify the analysis as much as possible,
we choose the integers $m$ and $n$ so that $\th_k^\ast$
reduces to the value of $\th_k$ defined for hydrogen in
Eq.\ \rf{etatheta}.
This imposes
\beq
\fr m n =
\fr {3 p (\la \nu - \mu)} {4 q \nu}
\quad\quad {\rm (mod \, 1)}
\quad ,
\label{mnstar}
\eeq
where the right-hand side is understood to be fully reduced.
With this choice,
the phase $\th_k^\ast$ equals the phase
$\th_k$ for hydrogen given in
Eq.\ \rf{etatheta}.
It therefore has the same minimum period $l$
under shifts in the summation index $k$
as for hydrogen,
given in Eq.\ \rf{lvalues}.

Note that if $\mu = 0$,
the definition \rf{mnstar}
for $m/n$ reduces to the form in
Eq.\ \rf{mn} for hydrogen.
The additional shift for $\mu \ne 0$ comes
from the fractional part of ${\bar n}^\ast$
due to the quantum defect,
assuming the laser is on resonance.
We therefore see that the allowed times
$t_{\rm frac}^\ast$ are \it not \rm simply obtained
by the scaling transformations
$t_{\rm sr} \rightarrow t_{\rm sr}^\ast$ and
$t_{\rm rev} \rightarrow t_{\rm rev}^\ast$
in the definition of $t_{\rm frac}$.
In addition to the scaling,
the fraction $m/n$ is shifted by
an amount that depends on the
quantum defect.

We next consider the phase $\ph_k^\ast$ induced
by shifting the classical waves
$\psi_{\rm cl} ({\vec r},t)$ by an amount
\beq
\De t = \fr a b t_{\rm rev}^\ast + \fr c d T_{\rm cl}^\ast
\quad ,
\eeq
where $a$, $b$ and $c$, $d$
are pairs of relatively prime integers.
The phase generated is
\beq
\ph_k^\ast = \fr {2 {\bar n}^\ast a} {3 b} k
+ \fr c d k
\quad .
\label{phikstar}
\eeq

Substituting the definition of
${\bar n}^\ast$ given in
Eq.\ \rf{nbarstar},
we choose the integers $c$ and $d$ to satisfy
\beq
\fr {2 (\la \nu - \mu) a} {3 \nu b}
+ \fr c d = 0
\quad\quad {\rm (mod \, 1)}
\quad .
\label{cdcondstar}
\eeq
With this condition,
$\ph_k^\ast$ reduces to
the expression \rf{phi2} for the
phase $\ph_k$ in hydrogen.

Following the definitions in
Eq.\ \rf{reduce},
$\ph_k$ can be written as ${\al}/l k$,
where $\al$ and $l$ are relatively prime integers
as defined in Sec.\ IIIC.
The time shifts $\De t$
are then given by
\beq
\De t = \fr {\al} l T_{\rm cl}^\ast
\quad .
\eeq
The minimum period of $\ph_k^\ast$ under shifts in the
summation index $k$ is again $l$,
which matches the period of $\th_k^\ast$.

Since the phases induced by the shifts $\De t$ in
$\psi_{\rm cl} ({\vec r},t)$
have the same period as the higher-order
contributions to the time-dependent phase in
$\Psi ({\vec r},t)$ at the times $t_{\rm frac}^\ast$,
we may use the set
$\psi_{\rm cl} ({\vec r},t_{\rm frac}^\ast + \fr {s \al} l
T_{\rm cl}^\ast)$
with $s = 0, 1, \ldots, l-1$
as a basis for an expansion of $\Psi ({\vec r},t_{\rm frac}^\ast)$,
and we obtain Eq.\ \rf{expansstar}.
Since the phases $\th_k^\ast$ and $\ph_k^\ast$
have the same formal structure as those in hydrogen,
the coefficients $b_s$
may be taken to have the same form as in Eq.\ \rf{bs}.
The proof that this expansion is valid follows as before,
using Eq.\ \rf{sum}.

The allowed values of $q$ in the
definition \rf{tfracstar} of $t_{\rm frac}^\ast$
are restricted to multiples of three.
The proof follows that in hydrogen.
Since the phase $\ph_k^\ast$ reduces to $\ph_k$ for hydrogen,
and since Eq.\ \rf{reduce} holds,
the quantity ${l N}/ 3$ must again be an integer.
This is true only if $l$ and hence $q$
are multiples of three.

We next prove that the wave packet and absolute square
of the autocorrelation function are periodic for times
near
$t_{\rm frac}^\ast$
and we determine the periodicity
$T_{\rm frac}^\ast$.
The procedure is similar to the corresponding proofs for
hydrogen.

Consider the wave function $\Psi ({\vec r},t_{\rm frac}^\ast)$
under a shift in time
$t_{\rm frac}^\ast \rightarrow t_{\rm frac}^\ast
+ \fr e f t_{\rm rev}^\ast$.
It can be shown that the minimum shift that leads to periodic
behavior is $e/f = 3/q$.
The wave function at this
shifted time can be written as an expansion in
$\psi_{\rm cl}$ with coefficients $b_s^\prime$ given in
Eq.\ \rf{bsprime}.
In parallel with the hydrogenic case,
for $p =1$
we find the relation
$\vert b_s^\prime \vert = \vert b_{(s+x)_{{\rm mod} \, l}} \vert$
for some integer $x$ that can be shown to exist.

Taking the squared modulus of the wave function
and using the assumption
that the subsidiary waves do not overlap,
we deduce
\beq
\vert \Psi ({\vec r},t_{\rm frac}^\ast
 + T_{\rm frac}^\ast) \vert^2
 = \vert \Psi ({\vec r},t_{\rm frac}^\ast) \vert^2
\quad ,
\label{periodicwavestar}
\eeq
with the period
$T_{\rm frac}^\ast = \fr 3 q t_{\rm rev}^\ast
- \fr u v T_{\rm cl}^\ast$.
The integers $u$ and $v$ satisfy a relation
that depends on the fractional part of the quantum defect:
\beq
\fr u v = \fr {2 (\et + \la) - 3} q
- \fr {2 \mu} {q \nu}
\quad\quad {\rm (mod \, 1)}
\quad .
\label{uvstar}
\eeq
Equation \rf{periodicwavestar}
shows that near the times $t_{\rm frac}^\ast$
the wave packet is periodic with period
$T_{\rm frac}^\ast$.

Using a similar technique,
we have proved that the square of the autocorrelation
function is also periodic at the times $t_{\rm frac}^\ast$,
with the same period $T_{\rm frac}^\ast$.

\vfill
\newpage

{\bf\noindent REFERENCES}
\vglue 0.4cm

\vfill\eject

\baselineskip=16pt
{\bf\noindent FIGURE CAPTIONS}
\vglue 0.4cm

\begin{description}

\item[{\rm Fig.\ 1:}]
The absolute square of the autocorrelation function for a
hydrogenic
Rydberg wave packet with ${\bar n} = 319$ and $\si = 2.5$
is plotted as a function of time in microseconds.
The plots span times of order of the
revival time $t_{\rm rev} \simeq 1.05$ $\mu$sec.
(a) $t=0$ to $t=0.6$ $\mu$sec,
(b) $t=0.6$ to $t=1.2$ $\mu$sec.

\item[{\rm Fig.\ 2:}]
The absolute square of the autocorrelation function for a
hydrogenic
Rydberg wave packet with ${\bar n} = 319$ and $\si = 2.5$
is plotted as a function of time in microseconds.
(a) $t=0$ to $t=15$ $\mu$sec,
(b) $t=15$ to $t=30$ $\mu$sec,
(c) $t=30$ to $t=45$ $\mu$sec.

\item[{\rm Fig.\ 3:}]
Unnormalized circular wave packets for hydrogen
with ${\bar n} = 319$ and $\si = 2.5$.
Cross-sectional slices of the wave packet in the plane
of the orbit and for $r = \left< r \right> =
\fr 1 2 {\bar n} (2{\bar n} + 1)$ are plotted as
a function of the azimuthal angle $\phi$ in radians.
(a) $t=0$,
(b) $t = t_{\rm rev}$,
(c) $t = \fr 1 {6} t_{\rm sr} - \fr 3 8 t_{\rm rev}$,
(d) $t = \fr 1 {6} t_{\rm sr} + \fr 1 8 t_{\rm rev}
- \fr 5 6 T_{\rm cl}$,
(e) $t = \fr 1 {9} t_{\rm sr} - \fr 1 4 t_{\rm rev}$,
(f) $t = \fr 1 {9} t_{\rm sr} + \fr {1} {12} t_{\rm rev}
- \fr 8 9 T_{\rm cl}$,
(g) $t = \fr 1 {12} t_{\rm sr} - \fr {3} {16} t_{\rm rev}$,
(h) $t = \fr 1 {12} t_{\rm sr} + \fr {1} {16} t_{\rm rev}
- \fr 5 {12} T_{\rm cl}$,
(i) $t = \fr 1 {15} t_{\rm sr} - \fr 3 {20} t_{\rm rev}$,
(j) $t = \fr 1 {15} t_{\rm sr} + \fr {1} {20} t_{\rm rev}
- \fr {11} {15} T_{\rm cl}$,
(k) $t = \fr 1 {18} t_{\rm sr} - \fr 1 8 t_{\rm rev}$,
(l) $t = \fr 1 {18} t_{\rm sr} + \fr {1} {24} t_{\rm rev}
- \fr {17} {18} T_{\rm cl}$.

\item[{\rm Fig.\ 4:}]
Unnormalized radial wave packets for hydrogen
with ${\bar n} = 36$ and $\si = 1.5$.
The radial probability density is plotted as
a function of $r$ in atomic units.
(a) $t=0$,
(b) $t = \fr 1 2 T_{\rm cl}$,
(c) $t = t_{\rm rev} + \fr 1 2 T_{\rm cl}$,
(d) $t = \fr 1 {6} t_{\rm sr} - \fr 1 6 T_{\rm cl}$,
(e) $t = \fr 1 {6} t_{\rm sr} + \fr 1 2 t_{\rm rev}
- \fr 2 3 T_{\rm cl}$.

\item[{\rm Fig.\ 5:}]
The absolute square of the autocorrelation function for a
Rydberg wave packet for rubidium
with ${\bar n} = 36$ and $\si = 1.5$
is plotted as a function of time in nanoseconds.
The p-state quantum defect is $\de (1) = 2.65$.

\item[{\rm Fig.\ 6:}]
Unnormalized radial wave packets for rubidium
with ${\bar n} = 36$ and $\si = 1.5$.
The p-state quantum defect is $\de (1) = 2.65$.
The radial probability density is plotted as
a function of $r$ in atomic units.
(a) $t=0$,
(b) $t = \fr 1 2 T_{\rm cl}^{\ast}$,
(c) $t = t_{\rm rev}^{\ast} + \fr {4} {15} T_{\rm cl}^{\ast}$,
(d) $t = \fr 1 {6} t_{\rm sr}^{\ast}
  - \fr {27} {160} t_{\rm rev}^{\ast}
  + \fr 2 5 T_{\rm cl}^{\ast}$,
(e) $t = \fr 1 {6} t_{\rm sr}^{\ast}
  + \fr {53} {160} t_{\rm rev}^{\ast}
  - \fr {13} {60} T_{\rm cl}^{\ast}$.

\end{description}

\vfill\eject
\end{document}